\documentclass{ws-mpla}
\usepackage[super]{cite}
\usepackage{amsmath,amssymb}
\usepackage{graphicx}
\usepackage{mathtools,xparse,MnSymbol}
\usepackage{multirow}
\usepackage{stackrel}
\begin{document}

\def\bra#1{\langle #1 |}
\def\ket#1{| #1 \rangle}
\def\inner#1#2{\langle #1 | #2 \rangle}
\def\brac#1{\llangle #1 \|}
\def\ketc#1{\| #1 \rrangle}
\def\innerc#1#2{\llangle #1 \| #2 \rrangle}
\def\Lbra#1{\bigl\langle #1 \bigr|}
\def\Lket#1{\bigl| #1 \bigr\rangle}
\def\Linner#1#2{\bigl\langle #1 \big| #2 \bigr\rangle}
\def\Lbrac#1{\bigl\llangle #1 \bigr\|}
\def\Lketc#1{\bigl\| #1 \bigr\rrangle}
\def\Linnerc#1#2{\bigl\langle #1 \big| #2 \bigr\rangle}

\def\lg{\mathrel{\lower2.5pt\vbox{\lineskip=0pt\baselineskip=0pt
           \hbox{$<$}\hbox{$>$}}}}

\newcommand{\drawsquare}[2]{\hbox{%
\rule{#2pt}{#1pt}\hskip-#2pt
\rule{#1pt}{#2pt}\hskip-#1pt
\rule[#1pt]{#1pt}{#2pt}}\rule[#1pt]{#2pt}{#2pt}\hskip-#2pt
\rule{#2pt}{#1pt}}
\newcommand{\vev}[1]{ \langle {#1} \rangle }

\makeatletter
\newcommand{\subalign}[1]{%
  \vcenter{%
    \Let@ \restore@math@cr \default@tag
    \baselineskip\fontdimen10 \scriptfont\tw@
    \advance\baselineskip\fontdimen12 \scriptfont\tw@
    \lineskip\thr@@\fontdimen8 \scriptfont\thr@@
    \lineskiplimit\lineskip
    \ialign{\hfil$\m@th\scriptstyle##$&$\m@th\scriptstyle{}##$\hfil\crcr
      #1\crcr
    }%
  }%
}
\makeatother

\markboth{Yasunori Nomura}
{From the Black Hole Conundrum to the Structure of Quantum Gravity}

\catchline{}{}{}{}{}

\title{FROM THE BLACK HOLE CONUNDRUM TO THE STRUCTURE OF QUANTUM GRAVITY}

\author{\footnotesize YASUNORI NOMURA}

\address{Berkeley Center for Theoretical Physics, Department of Physics, University of California\\ and Theoretical Physics Group, Lawrence Berkeley National Laboratory,\\ Berkeley, CA 94720, USA\\
 Kavli Institute for the Physics and Mathematics of the Universe (WPI), UTIAS,\\ The University of Tokyo, Kashiwa, Chiba 277-8583, Japan\\
ynomura@berkeley.edu}

\maketitle


\begin{abstract}
We portray the structure of quantum gravity emerging from recent progress in understanding the quantum mechanics of an evaporating black hole. Quantum gravity admits two different descriptions, based on Euclidean gravitational path integral and a unitarily evolving holographic quantum system, which appear to present vastly different pictures under the existence of a black hole. Nevertheless, these two descriptions are physically equivalent. Various issues of black hole physics---including the existence of the interior, unitarity of the evolution, the puzzle of too large interior volume, and the ensemble nature seen in certain calculations---are addressed very differently in the two descriptions, still leading to the same physical conclusions. The perspective of quantum gravity developed here is expected to have broader implications beyond black hole physics, especially for the cosmology of the eternally inflating multiverse.
\end{abstract}


\section{Introduction}

Having a complete quantum theory of gravity has long been a major goal of theoretical physics.
This is because a naive merger of quantum mechanics and general relativity---though it works in certain limited regimes---suffers from major theoretical problems.
These problems can be divided into two categories.
One is the loss of predictivity for processes involving energies larger than the Planck scale, resulting from uncontrollable quantum corrections.
This problem is largely addressed with the knowledge we already have about string theory.
While we do not know the full structure of the theory, evidence suggests that we are on a right track.~\cite{Polchinski:1998rq}

The other is a fundamental structural problem, known broadly as the black hole information paradox.
In 1974, Hawking discovered that a black hole radiates at the quantum level, despite the fact that it only absorbs particles at the classical level.~\cite{Hawking:1974sw}
Together with the earlier suggestion by Bekenstein that a black hole has an entropy proportional to its horizon area,~\cite{Bekenstein:1973ur} this established the thermodynamics of black holes.
This great discovery, however, led to a peculiar conclusion.
Hawking's calculation seemed to indicate that information is lost in the process of formation and evaporation of a black hole.~\cite{Hawking:1976ra}
In other words, unitarity---one of the fundamental principles of quantum mechanics---did not seem to be preserved in such a process.

A major breakthrough regarding the issue occurred in 1997 when Maldacena discovered the anti-de~Sitter (AdS)/conformal field theory (CFT) correspondence,~\cite{Maldacena:1997re} a concrete realization of the broader idea called holography.~\cite{tHooft:1993dmi,Susskind:1994vu,Bousso:2002ju}
This correspondence asserts that physics occurring in gravitational spacetime with asymptotically AdS boundary conditions (called the bulk) is equivalent to that of a CFT defined in a lower-dimensional {\it non}-gravitational spacetime (called the boundary).
This allows us to map the process of black hole formation and evaporation in the bulk to a process of the CFT on the boundary, which is manifestly unitary.
This indicates that the formation and evaporation of a black hole must preserve information.

\begin{figure}[h!]
\centerline{\includegraphics[height=2.5in,trim = {0 0 0 1.5cm},clip]{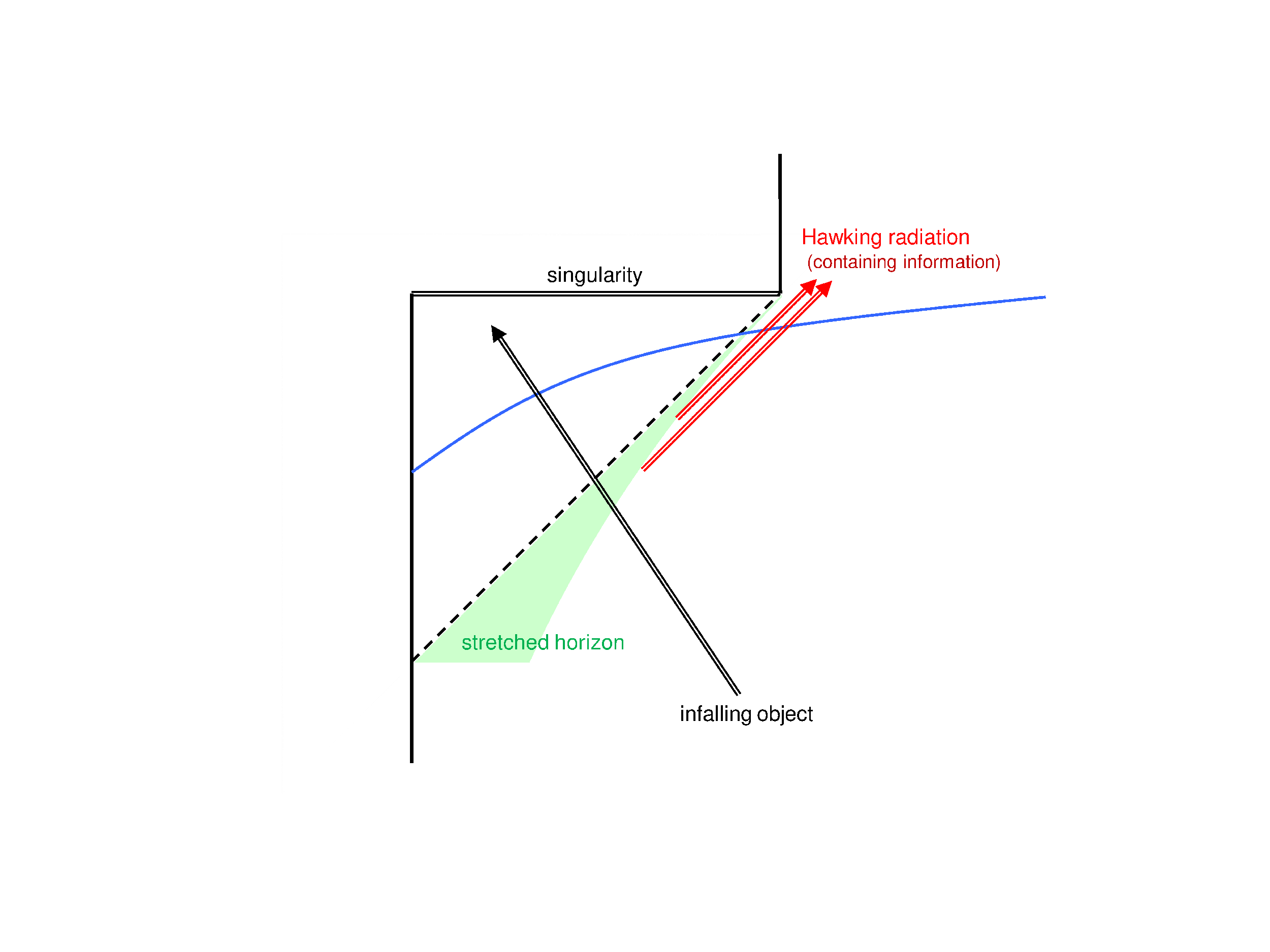}}
\caption{If Hawking radiation contains information about a fallen object, a late equal-time slice (blue curve) appears to have duplicate information (Hawking radiation and the object itself), which contradicts the no-cloning theorem of quantum mechanics.}
\label{fig:zerox}
\end{figure}

A problem in this picture is that it seems to be at odds with the existence of the black hole interior, a prediction of the equivalence principle of general relativity (and which Hawking's calculation assumed).
One way to see this is the following.
If the black hole evolution is indeed unitary, then the information about an object that falls into the black hole will be sent back later in Hawking radiation, which occurs well before the final stage of the evaporation.~\cite{Page:1993wv,Hayden:2007cs}
This implies that there exists a late equal-time hypersurface that goes through both the interior and exterior of the black hole on which the information about the object lies both in Hawking radiation and in the interior; see Fig.~\ref{fig:zerox}.
This contradicts the no-cloning theorem of quantum mechanics,~\cite{Wootters:1982zz} which states that quantum information cannot be faithfully copied.
An important point is that this hypersurface (called a nice slice) is totally legitimate from the point of view of semiclassical gravity; for example, all the curvature invariants associated with it are much smaller than the Planck scale.~\cite{Lowe:1995ac}
The failure, therefore, cannot be attributed to the unknown ultraviolet (UV) physics at the Planck scale, at least directly.

A suggested idea addressing this issue is called black hole complementarity, which claims that the problem is only academic because no physical observer can access both of the duplicated information even in principle.~\cite{Susskind:1993if,Susskind:1993mu}
The precise implementation of this idea, however, has not been clear, and there is even an argument (called the firewall paradox) claiming that the idea can in fact never be implemented.~\cite{Almheiri:2012rt,Almheiri:2013hfa,Marolf:2013dba}~%
\footnote{Nevertheless, a version of complementarity is realized in our final picture, as we will see below.}

The discussion above shows that the problem of black hole information is in fact the problem of reconciling unitarity with the existence of the interior.
A gist of this article is to elucidate how this issue has been addressed by recent theoretical progress.
The basic picture is that quantum mechanics (at least, in its current formulations) allows for representing only one of the unitarity and interior manifest, and the other---whichever not chosen---arises as a consequence of the dynamics of the theory in rather subtle ways.~\cite{Langhoff:2020jqa}
Although these two descriptions appear very different, they are in fact physically equivalent due to large nonperturbative gauge redundancies of a gravitational theory, which are much larger than the standard diffeomorphism~\cite{Marolf:2020xie,McNamara:2020uza} and relate even spaces with different topologies~\cite{Marolf:2020xie,Jafferis:2017tiu}.

Quantum descriptions of black holes have recently been advanced by a number of authors, most relevantly in Refs.~\refcite{Penington:2019npb,Almheiri:2019psf,Almheiri:2019hni,Penington:2019kki,Almheiri:2019qdq,Hashimoto:2020cas,Hartman:2020swn} for a description that manifestly has the interior and in Refs.~\refcite{Papadodimas:2012aq,Papadodimas:2013jku,Papadodimas:2015jra,Maldacena:2013xja,Nomura:2018kia,Nomura:2019qps,Nomura:2020ska} for a description that is manifestly unitary.
(For more complete references for the two descriptions, see Refs.~\refcite{Almheiri:2020cfm} and \refcite{Nomura:2020ska}, respectively.)
It is a curious fact that these developments have been made without a clear realization that they are dealing with two different descriptions.
Another purpose of this paper is to explicate how the two descriptions accommodate various features of a black hole in different---though physically equivalent---manners.
This illuminates how quantum gravity works under the environment of strong gravity, especially when the system develops a horizon.

The organization of this paper is the following.
In Section~\ref{sec:dual}, we give an overview of how the two descriptions work and list the problems of black hole physics that we want to address with them.
In the following two sections, we detail the two descriptions:\ Section~\ref{sec:global} for the description keeping the interior manifest and Section~\ref{sec:unitary} for that making unitarity manifest.
In Section~\ref{sec:beyond}, we briefly comment on implications of our analyses beyond the context of black hole physics.
Concluding discussion is given in Section~\ref{sec:discuss}.
Throughout the paper, we adopt natural units $c = \hbar = 1$.

\section{Quantum Gravity and Legendre Duality}
\label{sec:dual}

As discussed in the introduction, quantum gravity allows for two different descriptions of a system.
Under certain circumstances, especially when a horizon exists, the two descriptions give dramatically different representations of the same physics.~\cite{Langhoff:2020jqa,Nomura:2020ska,Harlow:2020bee}
The equivalence of these descriptions is guaranteed by large nonperturbative gauge redundancies of a gravitational theory.~\cite{Marolf:2020xie,McNamara:2020uza,Jafferis:2017tiu}
In this section, we discuss the defining characteristics of the two descriptions.
We also list a number of issues that will be addressed in the following sections, indeed rather differently in the two descriptions.

As will become clearer below, the two descriptions can be naturally associated with two formulations of quantum mechanics:\ the path integral and canonical formalisms.
Since these formalisms go well with the Lagrangian and Hamiltonian approaches, respectively, and they are related by the Legendre transformation, we might call the existence and equivalence of the two descriptions Legendre duality.
We now overview each of these descriptions in turn.

\subsection*{Global spacetime description}

This description begins with the global spacetime picture of general relativity.
Quantization surfaces can be taken to be global equal-time hypersurfaces in general relativity.
In the context of black hole physics, these can be nice slices depicted in Fig.~\ref{fig:zerox}, which extend smoothly to both the exterior and interior of the black hole~\cite{Lowe:1995ac,Giddings:2006sj,Rosabal:2020dlq}.
Recent progress has revealed that this description is extremely redundant, much more than what is suggested by the standard diffeomorphism.
In particular, many of the states that are orthogonal in semiclassical gravity have exponentially small overlaps at the level of full quantum gravity, and this dramatically reduces the number of independent states from that one naively expect based on intuition of semiclassical spacetime.

\begin{figure}[h!]
\centerline{\includegraphics[height=1.6in,trim={0 3.8cm 0 3.3cm},clip]{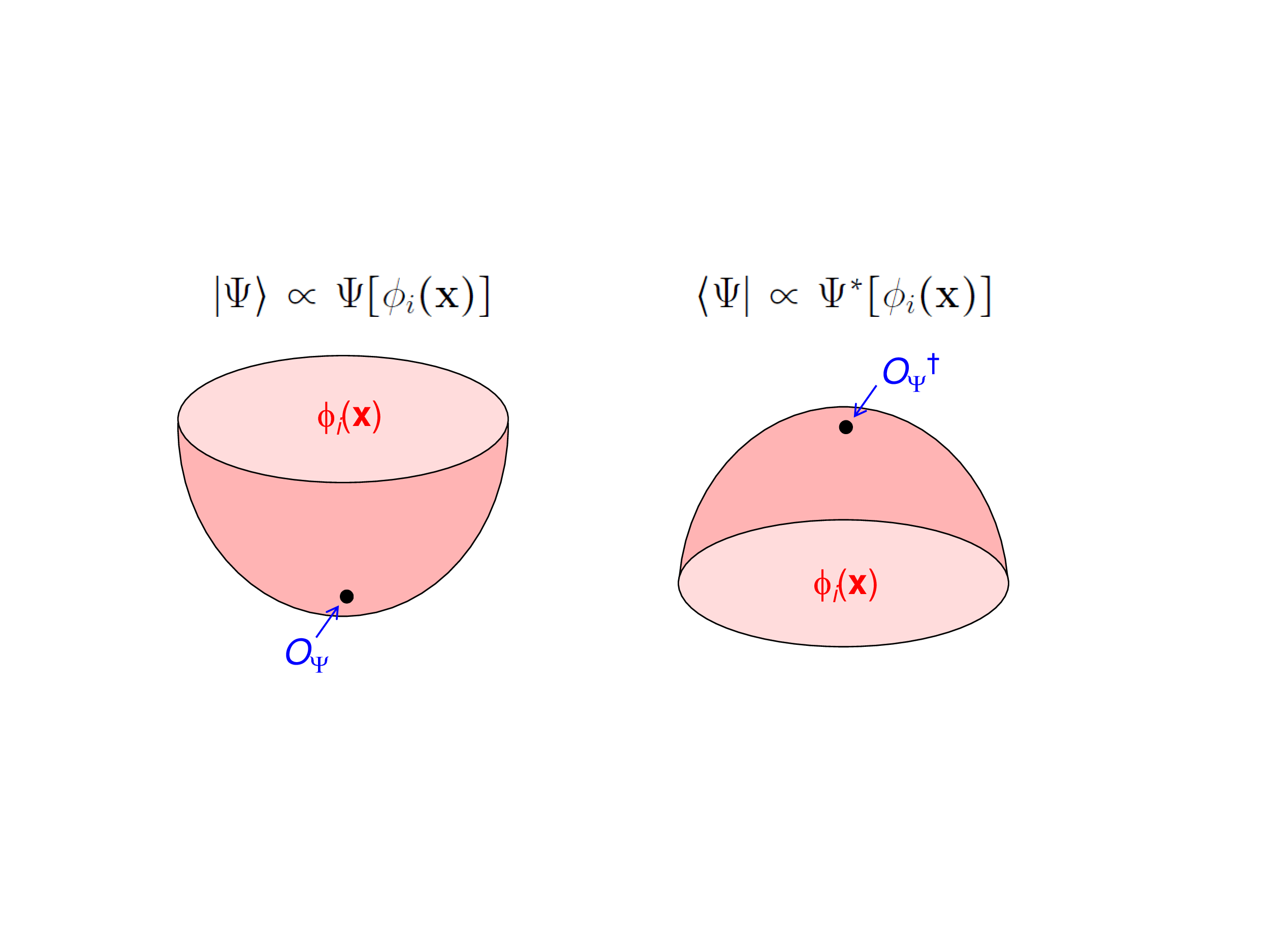}}
\caption{The ket vector $\ket{\Psi}$ is obtained by Euclidean path integral with the corresponding operator ${\cal O}_\Psi$ inserted in the past; in particular, the wavefunctional $\Psi[\phi_i({\bf x})]$ is given by performing the path integral for each fixed field configuration $\phi_i({\bf x})$ on spacelike hypersurface $S_0$ (left).
 Similarly, the conjugate vector $\bra{\Psi}$ is obtained by Euclidean path integral with operator ${\cal O}_\Psi^\dagger$ inserted in the future (right).
 Here, we have shown only the leading contributions.}
\label{fig:E-states}
\end{figure}

A natural platform for this description is Euclidean gravitational path integral,~\cite{Penington:2019kki,Almheiri:2019qdq,Harlow:2020bee,Padmanabhan:2019art} which treats the black hole as a quasi-static system.
In this method, a state (ket) $\ket{\Psi}$ is given as a functional $\Psi[\phi_i({\bf x})]$ of field configurations on an equal-time hypersurface $S_0$.
Here, ${\bf x}$ represents {\it spatial} coordinates on $S_0$, whose dimension need not be fixed throughout the state, and the index $i$ collectively denotes species.
The functional is determined by Euclidean gravitational path integral performed for each field configuration on $S_0$, with the operator ${\cal O}_\Psi$ corresponding to the state $\ket{\Psi}$ inserted in the past; see the left panel of Fig.~\ref{fig:E-states}.
Similarly, a conjugate state (bra) $\bra{\Psi}$ is given as a functional $\Psi^*[\phi_i({\bf x})]$ obtained by path integral with the operator ${\cal O}_\Psi^\dagger$ inserted in the future, as in the right panel of Fig.~\ref{fig:E-states}.

\begin{figure}[h!]
\centerline{\includegraphics[height=1.9in,trim={0 0 0 0.5cm},clip]{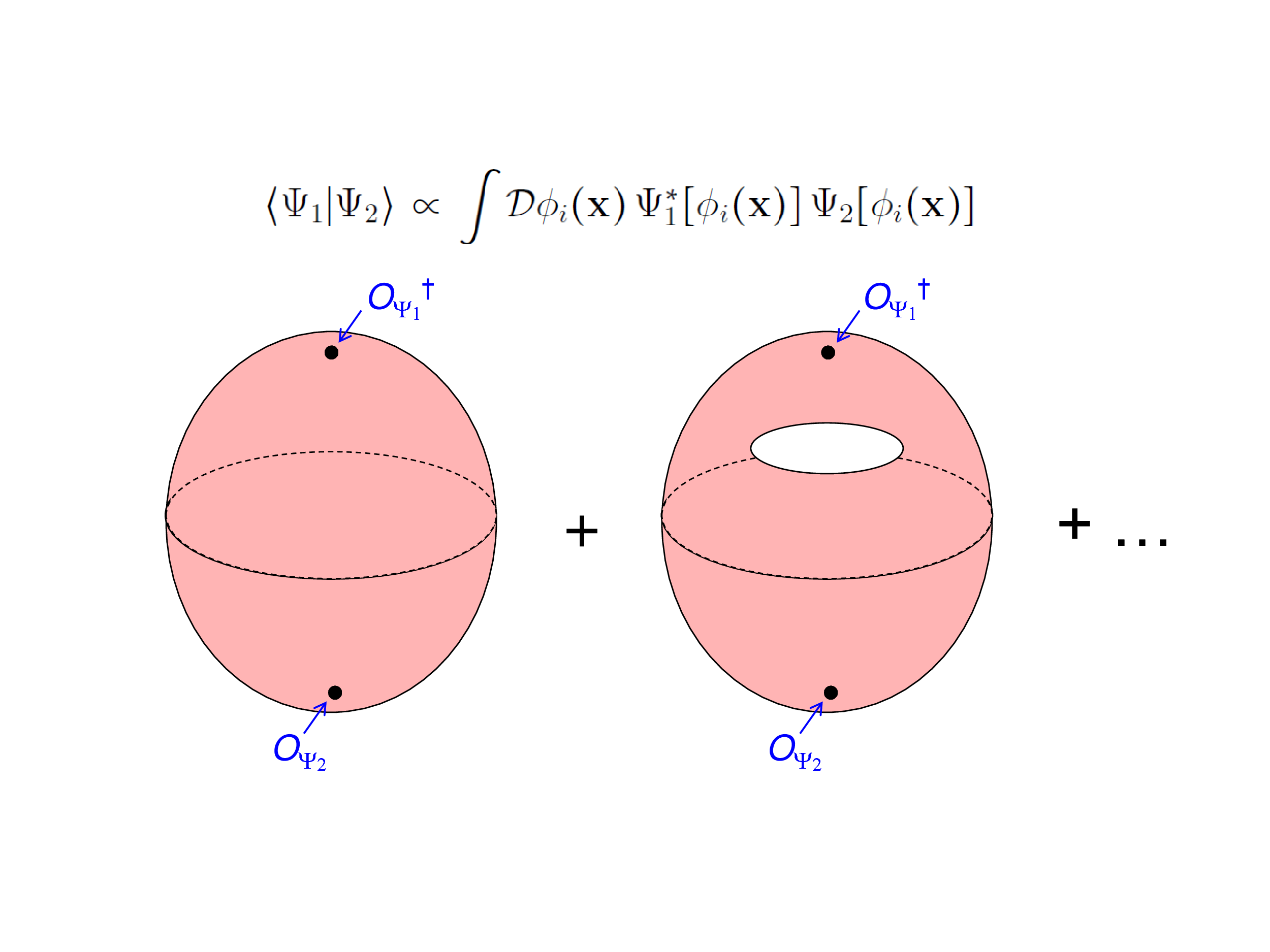}}
\caption{The inner product $\inner{\Psi_1}{\Psi_2}$ between two states $\ket{\Psi_1}$ and $\ket{\Psi_2}$ is given by a Euclidean path integral in which operators ${\cal O}_{\Psi_1}^\dagger$ and ${\cal O}_{\Psi_2}$ are inserted in the future and past.}
\label{fig:E-inner}
\end{figure}

The inner product of two states, $\inner{\Psi_1}{\Psi_2}$, is given by a path integral in which operators ${\cal O}_{\Psi_1}^\dagger$ and ${\cal O}_{\Psi_2}$ are inserted in the future and past, respectively.
An important point is that in performing this path integral, all possible ``appropriately smooth'' spacetimes must be integrated, including those with different topologies; see Fig.~\ref{fig:E-inner}.
The claim is that inclusion of all these spacetime histories amounts to incorporating the effect that states orthogonal at the semiclassical level may have nonzero overlaps.
Specifically, as the semiclassical inner product $\inner{\Psi_1}{\Psi_2}$ becomes smaller, the true inner product also becomes smaller; however, this decrease saturates at $\inner{\Psi_1}{\Psi_2} \sim e^{-S/2}$ in full quantum gravity, where $S$ is the number of degrees of freedom participating nontrivially in the inner product.~\cite{Papadodimas:2015jra,Maldacena:2001kr}
As we will see, this effect addresses various problems in black hole physics,~\cite{Langhoff:2020jqa,Chakravarty:2020wdm} including unitarity.~\cite{Penington:2019kki,Almheiri:2019qdq}

\subsection*{Unitary\,/\,holographic description}

This description---which we refer to as the unitary gauge description---makes unitarity manifest.
In the case of a system with a black hole, this corresponds to viewing the black hole from a distance.
In this view, an object falling toward the black hole never crosses the horizon at the classical level due to infinite time delay.
At the quantum level, the horizon is ``stretched'' to a timelike surface called the stretched horizon,~\cite{Susskind:1993if} on which the local (Tolman) Hawking temperature becomes the string scale.
An infalling object is then absorbed into the stretched horizon in a finite time, becoming a part of the black hole.

Since the intrinsic scale of the dynamics on the stretched horizon is the string scale, it cannot be described by a low energy theory.
It is thus consistent to assume that the dynamics there---and hence the process of Hawking emission---is unitary (since Hawking's conclusion based on the global spacetime picture obviously does not apply).~\cite{Susskind:1993if,tHooft:1990fkf}.
In other words, the degrees of freedom outside (and on) the stretched horizon comprise the entirety of the system.
Indeed, this is the description a boundary theory in holography naturally leads to.

A challenge in this description is to understand how a description based on near empty interior spacetime emerges.
In particular, we must understand why such a description applies only to the stretched horizon; after all, from the viewpoint of quantum information flow, the stretched horizon is not too different from the surface of regular material such as a piece of coal.
As we will see, special universal properties of the string scale dynamics are at play in answering these questions.~\cite{Langhoff:2020jqa,Nomura:2018kia,Nomura:2019qps,Nomura:2020ska}
In fact, the defining characteristic of a black hole in this description is not the appearance of an inescapable spacetime region {\it per se}, but rather the emergence of a surface having these dynamical properties.

\subsection*{Black hole conundrums}

In the next two sections, we will discuss the above two descriptions in detail in the context of black hole physics.
Before doing so, it is useful to list what aspects of black hole physics we want to understand using these descriptions.
The first has already been discussed---it is about the very definition of a black hole:
\begin{itemlist}
 \item \underline{\bf The defining characteristic of a black hole}\\
 In the global spacetime description, the defining characteristic of a black hole is as in general relativity:\ the appearance of a spacetime region from which nothing can escape.
 In the unitary gauge description, on the other hand, the formation of a black hole is signaled by the emergence of a surface (stretched horizon) possessing special dynamical properties, which we will discuss in more detail later.
\end{itemlist}
The other aspects are listed below with the overview of how they are addressed in each description.
\begin{itemlist}
 \item \underline{\bf The existence of the interior}\\
 In the global spacetime description, the existence of the interior is evident by construction (at the cost of obscuring unitarity).
 In the unitary gauge description, the black hole interior emerges effectively as a collective phenomenon of fundamental degrees of freedom, through the universal dynamics of the stretched horizon.~\cite{Nomura:2018kia,Nomura:2019qps,Nomura:2020ska}
 \item \underline{\bf Unitarity of the evolution}\\
 In the global spacetime description, unitarity is seen after taking into account nonperturbative effects of quantum gravity, described by topologically nontrivial configurations in Euclidean gravitational path integral.~\cite{Penington:2019kki,Almheiri:2019qdq}
 In the unitary gauge description, unitarity is manifest (as the name suggests) as implied by holography and AdS/CFT in particular.
 \item \underline{\bf Bekenstein-Hawking entropy}\\
 Depending on hypersurfaces one chooses, the interior of a black hole can have an ever increasing spatial volume~\cite{Christodoulou:2014yia,Christodoulou:2016tuu}, which does not seem to be consistent with the Bekenstein-Hawking entropy.
 This problem is addressed in the global spacetime description because naively independent semiclassical interior states are actually not independent~\cite{Langhoff:2020jqa,Chakravarty:2020wdm}, and in the unitary gauge description because an effective theory of the interior describes only a limited spacetime region.~\cite{Langhoff:2020jqa,Nomura:2018kia}
 \item \underline{\bf The ensemble nature}\\
 In the analysis of a $(1+1)$-dimensional model of black hole evaporation, Ref.~\refcite{Penington:2019kki} found a result that seemed to require the interpretation that Euclidean gravitational path integral computes the average of a quantity over some ensemble, raising the question of the origin and identity of such an ensemble.
In the unitary gauge description, this is understood to arise from an average over black hole microstates.~\cite{Langhoff:2020jqa}
In the case of a $(1+1)$-dimensional theory, this averaging comes from the fact that its boundary dual is an ensemble of 1-dimensional theories.~\cite{Saad:2019lba,Stanford:2019vob}
\end{itemlist}

\section{Global Spacetime Description {\it a la}\, Euclidean Quantum Gravity}
\label{sec:global}

In this section, we discuss the description based on the global spacetime picture.
As already stated, the starting point of this description is global spacetime of general relativity, so that equal-time hypersurfaces, on which states (in the language of the Schr\"{o}dinger picture) are defined, go through both the exterior and interior of the black hole.
The existence of the black hole interior is evident by construction.
A challenge in this description is to understand how the evolution of a black hole can be unitary (when it is viewed from a distance), despite Hawking's analysis.~\cite{Hawking:1976ra,Mathur:2009hf}

This issue has been addressed recently by Refs.~\refcite{Penington:2019npb,Almheiri:2019psf,Almheiri:2019hni,Penington:2019kki,Almheiri:2019qdq}, using technologies developed in the study of holography~\cite{Ryu:2006bv,Hubeny:2007xt,Faulkner:2013ana,Engelhardt:2014gca} which are related to Euclidean gravitational path integral.
(For other early contributions, see, e.g., Refs.~\refcite{Hashimoto:2020cas,Hartman:2020swn,Almheiri:2019yqk,Rozali:2019day,Chen:2019uhq,Almheiri:2019psy,Gautason:2020tmk,Hollowood:2020cou,Krishnan:2020oun} and those in Ref.~\refcite{Almheiri:2020cfm}.)
Below, we present the basic idea using the language of Euclidean gravitational path integral.
Strictly speaking, the validity of this picture has been demonstrated explicitly only in certain lower-dimensional setups,~\cite{Penington:2019kki,Almheiri:2019qdq} but we expect that it is applicable more generally.~\cite{Hartman:2020swn}

\begin{figure}[h!]
\centerline{\includegraphics[height=1.9in,trim={0 0 0 0.5cm},clip]{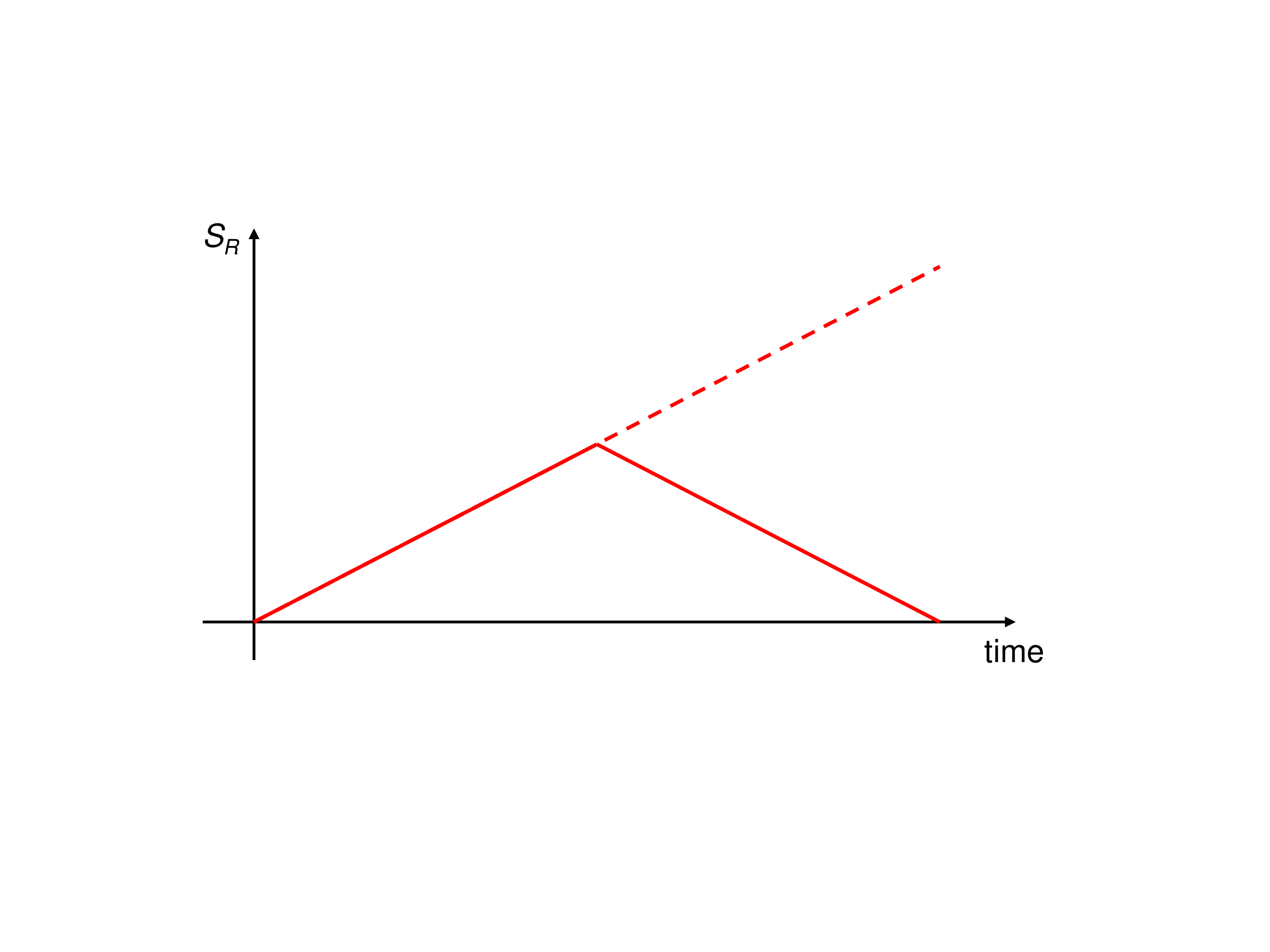}}
\caption{The von~Neumann entropy of Hawking radiation as a function of time.
 If the evolution of the black hole is unitary, then it follows the Page curve (depicted by the solid line).
 If the evolution is as indicated by Hawking's calculation leading to information loss, then it would behave as the thermal entropy (dashed line).}
\label{fig:Page}
\end{figure}

Our interest is the von~Neumann (fine-grained) entropy of the state of Hawking radiation as a function of time.
Suppose that a black hole is formed by collapsing matter which was in a pure state.
The von~Neumann entropy of Hawking radiation then increases initially reflecting the increasing entanglement between the radiation and the black hole generated by the Hawking emission process.
If the evolution of the black hole is unitary, however, this quantity must go back to zero at the end of the evolution, since the state of final Hawking radiation after the evaporation must be pure.
Assuming certain genericity conditions, the von~Neumann entropy follows what is called the Page curve,~\cite{Page:1993wv} depicted by the solid line in Fig~\ref{fig:Page}.
On the other hand, according to  Hawking's calculation, the von~Neumann entropy of radiation must follow its thermodynamic (coarse-grained) entropy depicted by the dashed line.
The final state of the radiation ends up with being mixed.

\begin{figure}[h!]
 \centerline{\includegraphics[height=2.4in,trim={0 0 0 0.8cm},clip]{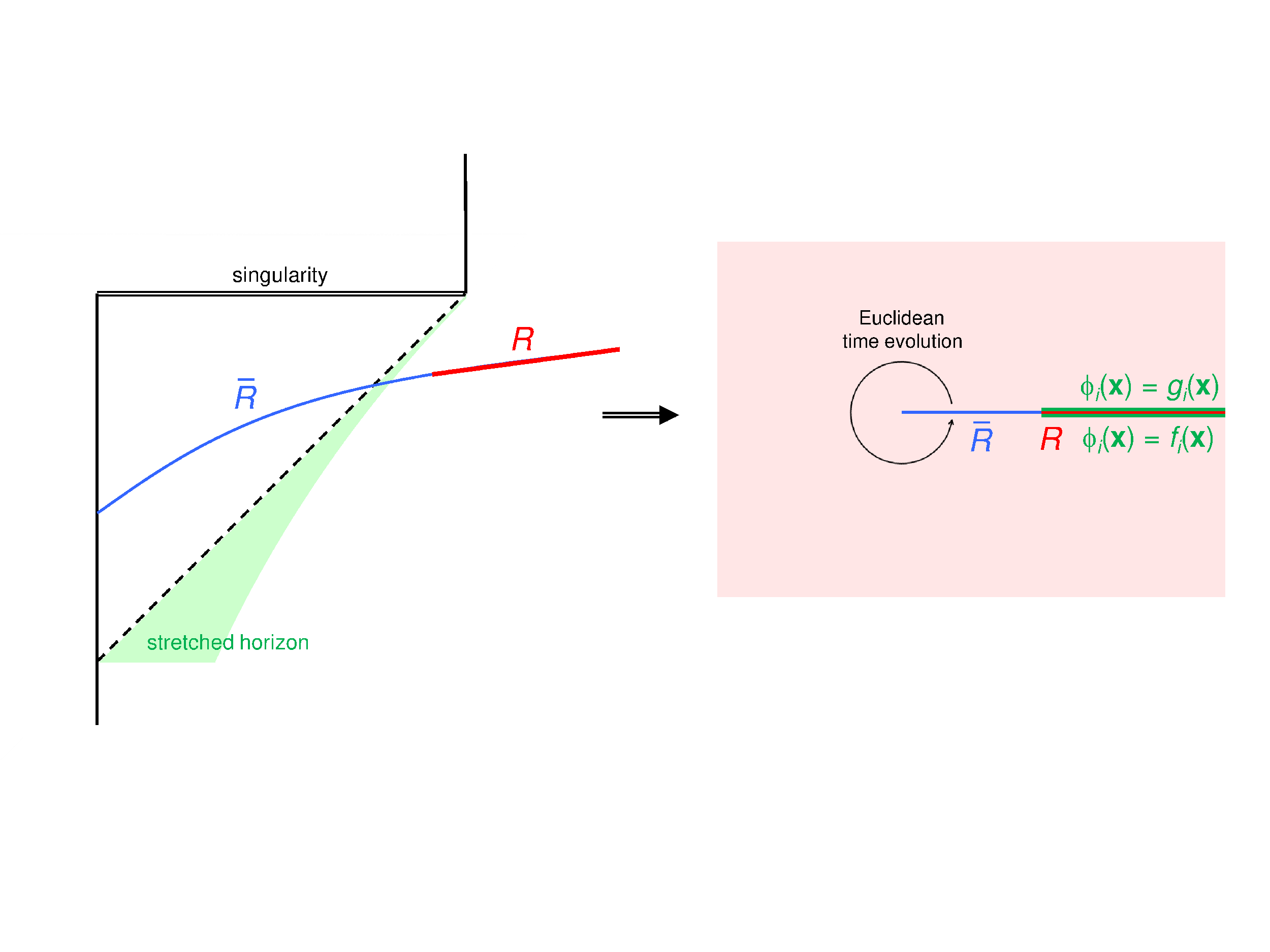}}
\caption{On a nice slice $\Sigma$ in global spacetime, one can consider a subregion $R$ on which emitted Hawking radiation resides (left).
 The reduced density matrix of $R$, obtained by tracing the complement $\bar{R}$ of a global state on $\Sigma$, is then given by Euclidean gravitational path integral in the form of a functional $\rho_R[f_i({\bf x}),g_i({\bf x})]$ of two field configurations $f_i({\bf x})$ and $g_i({\bf x})$ on $R$ (right).}
\label{fig:rho_R}
\end{figure}

To follow the time evolution of the von~Neumann entropy of radiation, let us consider a state $\ket{\Psi}$ on an equal-time hypersurface $\Sigma$, taken in global spacetime.
An example of such a hypersurface is a nice slice depicted in Fig.~\ref{fig:zerox}.
We focus on the reduced density matrix $\rho_R$ of a subregion $R$ outside the black hole (more precisely, outside the black hole zone region~\cite{Nomura:2018kia}) on which emitted radiation resides.
This density matrix is given by Euclidean gravitational path integral in the form of a functional of two sets of field configurations $f_i({\bf x})$ and $g_i({\bf x})$ on $\Sigma$ (see Fig.~\ref{fig:rho_R}):
\begin{equation}
  {\rm Tr}_{\bar{R}} \ket{\Psi} \bra{\Psi} = \rho_R[f_i({\bf x}),g_i({\bf x})],
\end{equation}
where $\bar{R}$ is the complement of $R$ on $\Sigma$.

\begin{figure}[h!]
 \centerline{\includegraphics[height=1.8in,trim={0 0 0 0},clip]{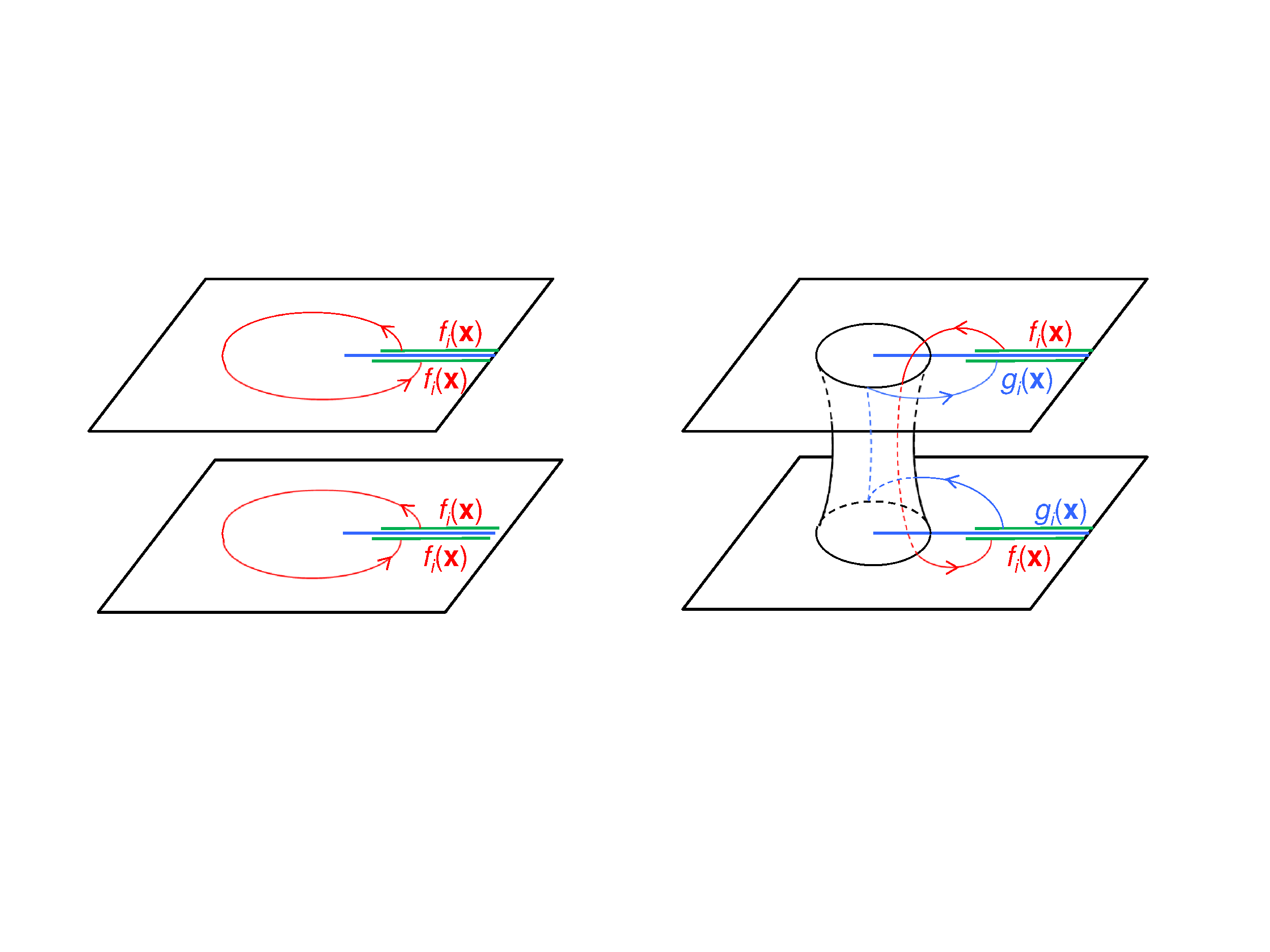}}
\caption{The trace of the square of the (unnormalized) reduced density matrix $\rho_R$ can be calculated by performing path integral on two copies of the original Euclidean spacetime with appropriate (branching) boundary conditions on $R$, depicted by green lines.
 The naive contribution (corresponding to Hawking's result of information loss) comes from the disconnected topology shown on the left.
 Unitarity of the black hole evolution is recovered by including the contribution from a connected topology (Euclidean/replica wormhole) shown on the right.
 Here, $f_i({\bf x})$ and $g_i({\bf x})$ must be integrated.}
\label{fig:rho-2}
\end{figure}

The von~Neumann entropy of $R$ is given by
\begin{equation}
  S_R = -{\rm Tr}[\rho_R \ln\rho_R],
\end{equation}
where we have assumed that $\rho_R$ is appropriately normalized:\ ${\rm Tr}[\rho_R] = 1$.
It is easiest to compute this as a limit:
\begin{equation}
   S_R = \lim_{n \rightarrow 1} \frac{1}{1-n} \ln {\rm Tr}[\rho_R^n],
\end{equation}
since this can avoid taking the matrix logarithm, $\ln\rho_R$.
The quantity ${\rm Tr}[\rho_R^n]$ can be computed by the so-called replica trick.
For $n =2$, for example, one can prepare two copies of the Euclidean spacetime in the right panel of Fig.~\ref{fig:rho_R} and glue them in such a way that the resulting path integral corresponds to ${\rm Tr}[\rho_R^2]$; see the left panel of Fig.~\ref{fig:rho-2}.%
\footnote{In fact, the quantity obtained by the path integral does not satisfy the normalization condition of $\rho_R$.
 The normalization must be imposed explicitly by $\rho_R \rightarrow \rho_R/{\rm Tr}[\rho_R]$ as we will do later.}

The discovery of Refs.~\refcite{Penington:2019kki,Almheiri:2019qdq} is that when the spacetime is dynamical, there are other contributions to the path integral preserving the boundary conditions on $R$, which have different topologies as depicted in the right panel of Fig.~\ref{fig:rho-2}.
While these contributions are suppressed exponentially in the classical action, they become important under certain circumstances, including later stages in a black hole evaporation process.

As an example, let us estimate the two contributions to ${\rm Tr}[\rho_R^2]$, given in Fig.~\ref{fig:rho-2}.
First, each ``sheet'' contributes $e^{S_{\rm bh}}$, so that the left and right diagrams give $e^{2S_{\rm bh}}$ and $e^{S_{\rm bh}}$, respectively.
Here, $S_{\rm bh}$ is the classical action, identified in this case as the entropy of the black hole.
(More generally, this contribution gives $e^{\chi S_{\rm bh}}$, where $\chi$ is the Euler number of the configuration.)
Next, taking field configurations $\phi_i({\bf x})$ on $R$ to be eigenstates of Euclidean time evolution, a nontrivial cycle on a sheet forces $\phi_i({\bf x})$ appearing along it to be the same, as depicted in the figure.
Denoting the number of possible field configurations $\phi_i({\bf x})$ by $e^{S_{\rm rad}}$, since the state on $R$ is that of radiation, their trace provides factors of $e^{S_{\rm rad}}$ and $e^{2S_{\rm rad}}$ in the left and right panels, respectively.

Finally, the density matrix $\rho_R$ must be normalized.
This is ensured by dividing the whole contribution by ${\rm Tr}[\rho_R]^2$, where ${\rm Tr}[\rho_R]$ is given by a single sheet without branch cuts and hence $e^{S_{\rm bh} + S_{\rm rad}}$.
Thus, the total contribution from the diagrams in Fig.~\ref{fig:rho-2} is given by
\begin{equation}
  \frac{{\rm Tr}[\rho_R^2]}{{\rm Tr}[\rho_R]^2} = \frac{e^{2S_{\rm bh} + S_{\rm rad}} + e^{S_{\rm bh} + 2S_{\rm rad}}}{(e^{S_{\rm bh} + S_{\rm rad}})^2} = \frac{1}{e^{S_{\rm rad}}} + \frac{1}{e^{S_{\rm bh}}},
\end{equation}
where the first and second terms in the last expression come from the left and right diagrams, respectively.
In an early stage of the black hole evolution, when the coarse-grained entropy of radiation $S_{\rm rad}$ is smaller than the black hole entropy $S_{\rm bh}$, the first term from the trivial, disconnected topology (left diagram) dominates.
However, at a later stage when $S_{\rm rad}$ becomes larger than $S_{\rm bh}$, the dominant contribution switches to come from the connected topology (right diagram).

This analysis can be generalized to higher powers of $\rho_R$, yielding
\begin{equation}
  \frac{{\rm Tr}[\rho_R^n]}{{\rm Tr}[\rho_R]^n} = \frac{1}{e^{(n-1)S_{\rm rad}}} + \frac{1}{e^{(n-1)S_{\rm bh}}}.
\end{equation}
This, therefore, gives by analytic continuation
\begin{equation}
  S_R = \lim_{n \rightarrow 1} \frac{1}{1-n} \ln \frac{{\rm Tr}[\rho_R^n]}{{\rm Tr}[\rho_R]^n} 
  = \begin{cases}
    S_{\rm rad} & \mbox{for } S_{\rm rad} < S_{\rm bh} \\
    S_{\rm bh}  & \mbox{for } S_{\rm rad} > S_{\rm bh},
  \end{cases}
\end{equation}
reproducing the Page curve in Fig.~\ref{fig:Page}.
Note that if we did not include topologically nontrivial (connected) contributions, then we would have obtained
\begin{equation}
  S_R = S_{\rm rad},
\label{eq:Hawking}
\end{equation}
regardless of the relative size of $S_{\rm rad}$ and $S_{\rm bh}$.
This is Hawking's result.~\cite{Hawking:1976ra}
In other words, unitarity is recovered because of the topologically nontrivial contributions to the Euclidean gravitational path integral, called Euclidean or replica wormholes.~\cite{Penington:2019kki,Almheiri:2019qdq}

There is a simple, intuitive way to understand the result described above.
Recall that Hawking's result in Eq.~(\ref{eq:Hawking}) arises because one of a Hawking pair created at the horizon falls into the black hole while the other escapes to ambient space, and successive occurrences of this process keep increasing entanglement between the black hole and radiation as $S_{\rm rad}$.
In semiclassical gravity, the $e^{S_{\rm rad}}$ states of the fallen Hawking modes involved in the entanglement are all independent.
In full quantum gravity, however, these semiclassically independent states develop overlaps of order $e^{-S_{\rm bh}/2}$ when the number of fallen degrees of freedom $S_{\rm rad}$ exceeds the black hole entropy $S_{\rm bh}$.
Because of these overlaps, the number of independent states for the fallen Hawking modes is in fact only $e^{S_{\rm bh}}$, consistently with the Bekenstein-Hawking entropy (the rest being null after diagonalization).
The entanglement entropy between the black hole and radiation, therefore, goes as $S_{\rm bh}$ for $S_{\rm rad} > S_{\rm bh}$.

The same mechanism also addresses the problem of infinitely large spatial volumes inside a black hole.~\cite{Langhoff:2020jqa}
While a large spatial volume at a late time described in Refs.~\refcite{Christodoulou:2014yia,Christodoulou:2016tuu} seems to be able to host much larger number of independent quantum states than that indicated by the Bekenstein-Hawking entropy, these seemingly independent states are in fact not independent at the level of full quantum gravity.
Similarly, the large spatial volume of the so-called Wheeler's bags of gold, peculiar solutions of Einstein's equation in which there is a newly born universe inside a black hole, does not violate the Bekenstein-Hawking entropy bound.~\cite{Chakravarty:2020wdm}
In the next section, we discuss how these aspects are manifested in the unitary gauge description.

\begin{figure}[h!]
\centerline{\includegraphics[height=1.6in,trim={0 0 0 0},clip]{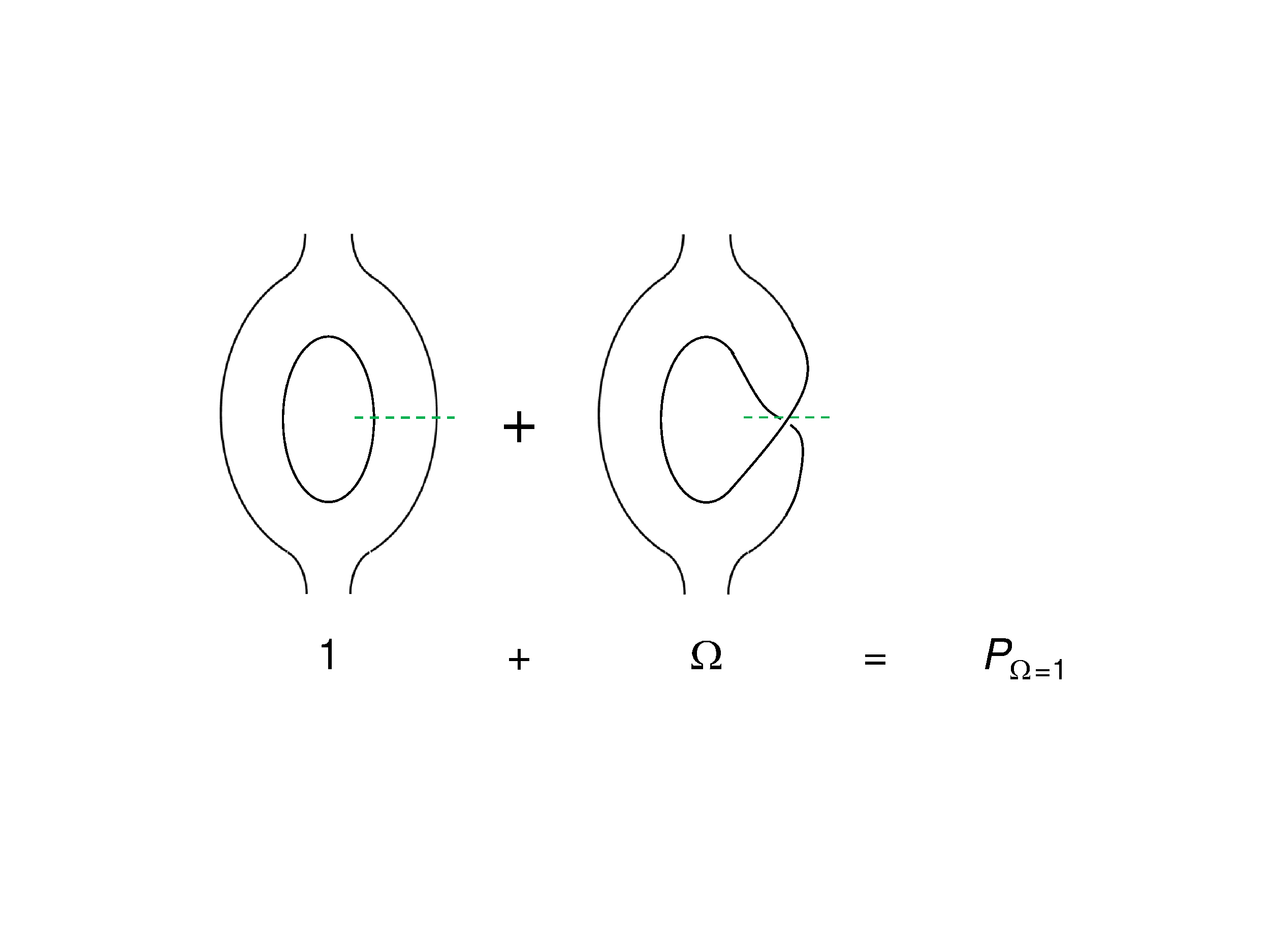}}
\caption{Summing a contribution and that obtained from it by performing an operation $\Omega$ (in this example, the parity transformation on a string worldsheet) in path integral amounts to projecting onto states that are invariant under $\Omega$.}
\label{fig:proj}
\end{figure}

One might think it odd that including more configurations in a path integral reduces the number of independent states, instead of increasing it.
This is, however, a general phenomenon.
Specifically, summing the contributions related by an operation $\Omega$ corresponds to projecting onto states singlet under the operation (see Fig.~\ref{fig:proj}).
This constitutes a reason behind the conjecture that the baby universe state (the state corresponding to a gravitational path integral without a boundary) is unique, ${\rm dim}{\cal H}_{\rm BU} = 1$,~\cite{Marolf:2020xie,McNamara:2020uza} which may have important implications for multiverse cosmology.~\cite{Nomura:2012zb}

Finally, we mention that explicit calculations along the lines described above in a $(1+1)$-dimensional gravitational theory found that the black hole interior states $\ket{\psi_A}$ satisfy~\cite{Penington:2019kki}
\begin{equation}
  \inner{\psi_A}{\psi_B} = \delta_{AB},
\qquad
  |\inner{\psi_A}{\psi_B}|^2 = \delta_{AB} + O\bigl(e^{-S_0}\bigr),
\label{eq:ens-lowerD}
\end{equation}
instead of
\begin{equation}
  \inner{\psi_A}{\psi_B} = \delta_{AB} + O\bigl(e^{-\frac{S_0}{2}}\bigr),
\qquad
  |\inner{\psi_A}{\psi_B}|^2 = \delta_{AB} \left\{ 1 + O\bigl(e^{-\frac{S_0}{2}}\bigr) \right\} + O\bigl(e^{-S_0}\bigr)
\label{eq:ens-higherD}
\end{equation}
as implied by the above discussion.
Based on the analysis in Ref.~\refcite{Langhoff:2020jqa}, we anticipate that this arises from coarse graining necessary to obtain the quasi-static picture of the system.
We will discuss this further in the next section using the unitary gauge description.

\section{Unitary Description {\it a la}\, Holography}
\label{sec:unitary}

In this section, we consider the unitary gauge description.
This description corresponds to a distant view of a black hole and hence corresponds to a boundary description in holography.
In this view, an object falling toward the black hole never reaches the horizon at the classical level because of infinite time delay caused by a diverging gravitational redshift.
At the quantum level, the horizon becomes a timelike surface called the stretched horizon,~\cite{Susskind:1993if} which is located where the local (Tolman) Hawking temperature becomes the string scale.
A falling object reaches this surface in a finite time.

\begin{figure}[h!]
\centerline{\includegraphics[height=1.9in,trim = {0 0 0 0},clip]{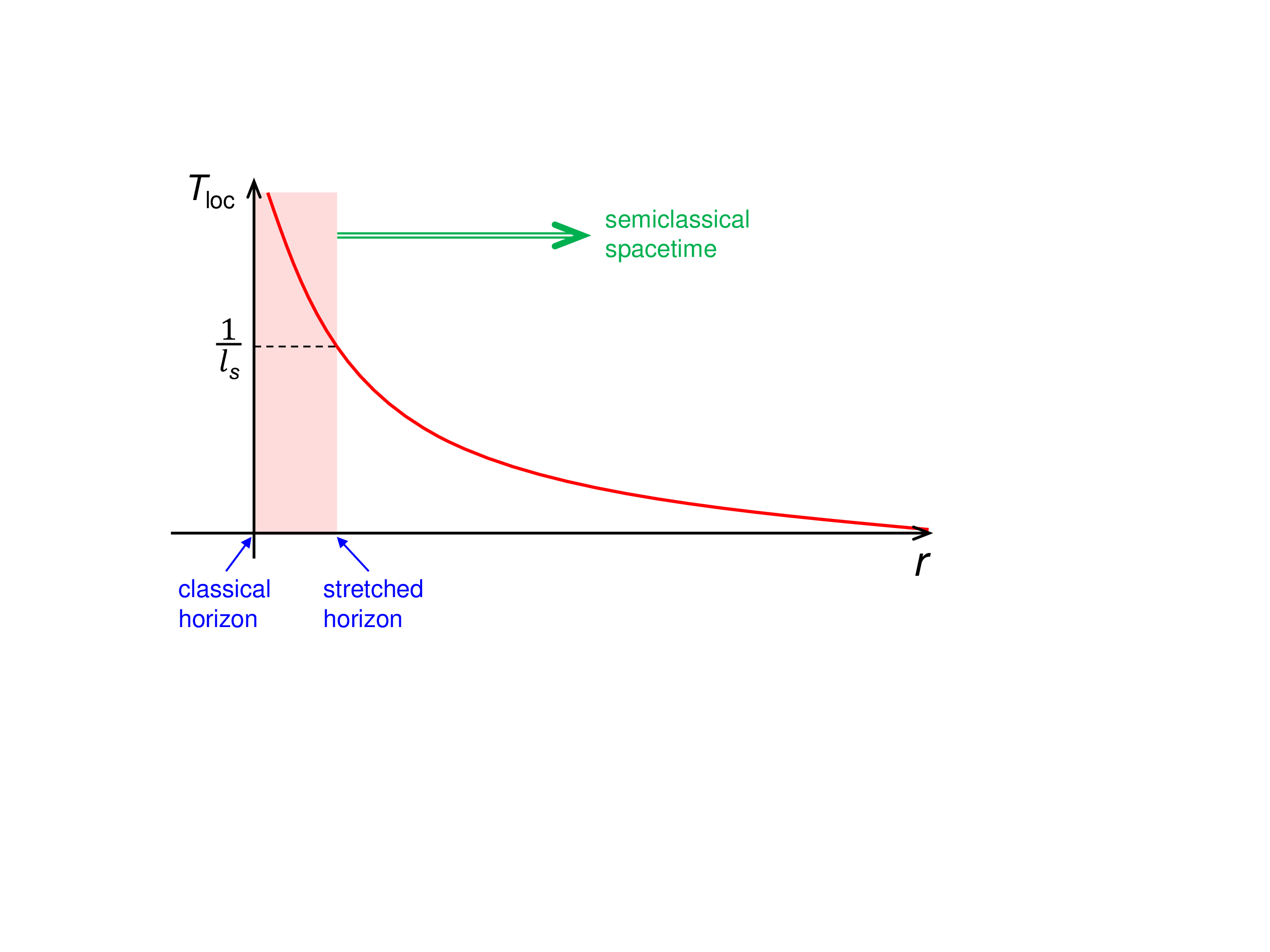}}
\caption{The local Hawking temperature $T_{\rm loc}$ as a function of the area radius $r$.
 The stretched horizon is located at $r = r_{\rm s}$, where $T_{\rm loc}$ becomes the string scale $1/l_{\rm s}$.
 The semiclassical spacetime picture is valid only for $r > r_{\rm s}$.
 The proper distance between the classical and stretched horizons is of order the string length $l_{\rm s}$.}
\label{fig:stretched}
\end{figure}

A basic tenet of this description is that the degrees of freedom outside the horizon comprise the entire system.%
\footnote{We may refer to this situation either as the entire degrees of freedom being ``outside and on the stretched horizon'' or simply as ``outside the stretched horizon.''
 In the rest of the paper, we adopt the latter for brevity.}
In other words, these degrees of freedom evolve unitarily under time evolution associated with a distant observer, or boundary time evolution in the language of holography.~\cite{tHooft:1990fkf,Susskind:1993if}
Hawking's analysis, which assumes the global semiclassical picture, obviously does not apply.
In the unitary gauge description, the scale of intrinsic dynamics (local Hawking temperature) is subject to a large gravitational blueshift near the horizon; for a static black hole
\begin{equation}
  T_{\rm loc}({\bf x}) = \frac{T_{\rm H}}{\sqrt{-g_{tt}({\bf x})}},
\label{eq:T_loc}
\end{equation}
where $T_{\rm H}$ is the Hawking temperature (as measured in the asymptotic region), and $t$ represents boundary time.
Since this scale reaches the string scale at the stretched horizon, physics there cannot be described by a low energy theory.
In fact, the classical spacetime description breaks down beyond this surface (see Fig.~\ref{fig:stretched}).
The starting point of the unitary gauge description is the assumption that this UV physics is unitary, as indeed implied by the AdS/CFT correspondence.

In the unitary gauge description, the stretched horizon thus behaves as a surface of regular material such as a piece of coal, as far as conservation of information is concerned.
The challenge in this description, therefore, is to understand how a picture based on near empty interior spacetime emerges.
In particular, we must understand why such a picture applies only to the stretched horizon.
This problem has been addressed in Refs.~\refcite{Papadodimas:2012aq,Papadodimas:2013jku,Papadodimas:2015jra,Maldacena:2013xja,Nomura:2018kia,Nomura:2019qps,Nomura:2020ska}, with the latter question answered in Refs.~\refcite{Nomura:2018kia,Nomura:2019qps}.
(For related work, see Refs.~\refcite{Verlinde:2012cy,Nomura:2012ex,Verlinde:2013qya,Nomura:2014woa,Papadodimas:2015xma,Nomura:2019dlz,Akal:2020ujg}; relations between these works are discussed in Ref.~\refcite{Nomura:2020ska}.)
Below, we discuss the description following Ref.~\refcite{Nomura:2020ska}, which provides the latest form of the framework.

\subsection*{Black hole as viewed from the exterior}

Suppose that there is a black hole of mass $M$ at a boundary time $t$.%
\footnote{A boundary time can be related to a bulk equal-time hypersurface through a gauge fixing procedure, for example by the ``holographic slice'' prescription of Refs.~\refcite{Nomura:2018kji,Murdia:2020iac}.}
For simplicity, we will focus on a spherically symmetric, non-near extremal black hole in asymptotically flat or AdS spacetime, although we expect that discussion below applies to any other non-extremal black hole.
Since we are dealing with an evaporating black hole, $M$ can be specified only up to the precision of
\begin{equation}
  \Delta = O(2\pi T_{\rm H}),
\end{equation}
determined by the uncertainty principle.
Below, we assume that the mass of a black hole is determined with this maximal precision.
A superposition of black holes of masses differing more than $\Delta$ can be treated in a straightforward manner.

In general, a black hole has a ``thermal atmosphere'' around it, which we refer to as the zone:
\begin{equation}
  r_{\rm s} \leq r \leq r_{\rm z},
\end{equation}
where $r$ is the area-radial coordinate, and $r_{\rm s}$ and $r_{\rm z}$ represent the locations of the stretched horizon and the edge of the zone, respectively.
The edge of the zone is determined by the gravitational potential generated by the black hole.
For a Schwarzschild black hole in four dimensions, for example, $r_{\rm s} - 2Ml_{\rm P}^2 = O(l_{\rm s}^2/Ml_{\rm P}^2)$ and $r_{\rm z} \approx 3Ml_{\rm P}^2$.
Here, $l_{\rm s}$ and $l_{\rm P}$ are the string and Planck lengths, respectively.
Analyses of more general black holes can be found in Ref.~\refcite{Langhoff:2020jqa}.

There are three classes of degrees of freedom associated with a black hole:\ hard modes, soft modes, and far modes (radiation)~\cite{Nomura:2018kia,Nomura:2019qps,Nomura:2020ska}.
First, there are modes in the zone whose characteristic frequencies (the frequencies and gaps among them) are sufficiently larger than $\Delta$.%
\footnote{When we refer to energy, frequency, and so on, we mean those as measured in the far, or asymptotic, region unless otherwise stated.}
These are ``coarse modes'' in the zone whose dynamics can be described by a semiclassical theory, and we call them the {\it hard modes}.
There are also modes in the zone whose frequencies are smaller than $\Delta$.
The dynamics of these modes, which we refer to as the {\it soft modes}, cannot be resolved by a semiclassical theory.
Different configurations of the soft modes give different microstates of the black hole.

The {\it far modes} are the modes located outside the zone, $r > r_{\rm z}$, whose dynamics can be described by semiclassical theory.
For our purposes, the relevant far modes are those that are entangled with the hard and/or soft modes.%
\footnote{Obviously, if we consider an object outside the zone which is falling toward the black hole, then modes describing the object would also be relevant.}
In our discussion below, we envision the simplest setup in which the relevant far modes consist of Hawking radiation emitted earlier from the black hole.
In general, they must involve all the modes entangled with the hard and soft modes, which may include degrees of freedom other than early Hawking radiation.
Including this effect, however, does not affect our discussion.

Let us begin by considering the situation in which there are no excitations beyond those directly associated with the existence of the black hole; namely, the system is in the semiclassical black hole vacuum.
This does not mean that hard modes, soft modes, and far modes are all in their ground states.
Due to entanglement between these modes and the energy constraint coming from the fact that the black hole has mass $M$, a black hole vacuum microstate, labeled by index $A$, takes the form~\cite{Nomura:2018kia,Nomura:2019qps,Nomura:2020ska}
\begin{equation}
  \ket{\Psi_{A,0}(M)} = \sum_n \sum_{i_n = 1}^{e^{S_{\rm bh}(M-E_n)}} 
    \sum_{a = 1}^{e^{S_{\rm rad}}} c^A_{n i_n a} \ket{\{ n_\alpha \}} 
    \ket{\psi^{(n)}_{i_n}} \ket{\phi_a},
\label{eq:sys-state}
\end{equation}
where the subscript $0$ on the left-hand side refers to the fact that the state is in the semiclassical vacuum.
$\ket{\{ n_\alpha \}}$, $\ket{\psi^{(n)}_{i_n}}$, and $\ket{\phi_a}$ are orthonormal states of the hard modes, soft modes, and far modes, respectively:
\begin{equation}
  \inner{\{ m_\alpha \}}{\{ n_\alpha \}} = \delta_{m n},
\qquad
  \inner{\psi^{(m)}_{i_m}}{\psi^{(n)}_{j_n}} = \delta_{m n} \delta_{i_m j_n},
\qquad
  \inner{\phi_a}{\phi_b} = \delta_{ab},
\label{eq:orthonorm}
\end{equation}
where $n \equiv \{ n_\alpha \}$ represents the set of all occupation numbers $n_\alpha$ ($\geq 0$) for the hard modes, which are labeled by $\alpha$ (collectively denoting the species, frequency, and angular-momentum quantum numbers),
$E_n$ is the energy of the hard mode state $\ket{\{ n_\alpha \}}$, and $S_{\rm bh}(E)$ is the Bekenstein-Hawking entropy density at energy $E$; see Ref.~\refcite{Langhoff:2020jqa} for a more detailed description of these states.

\begin{figure}[h!]
\centerline{\includegraphics[height=2.4in,trim = {0 0 0 0},clip]{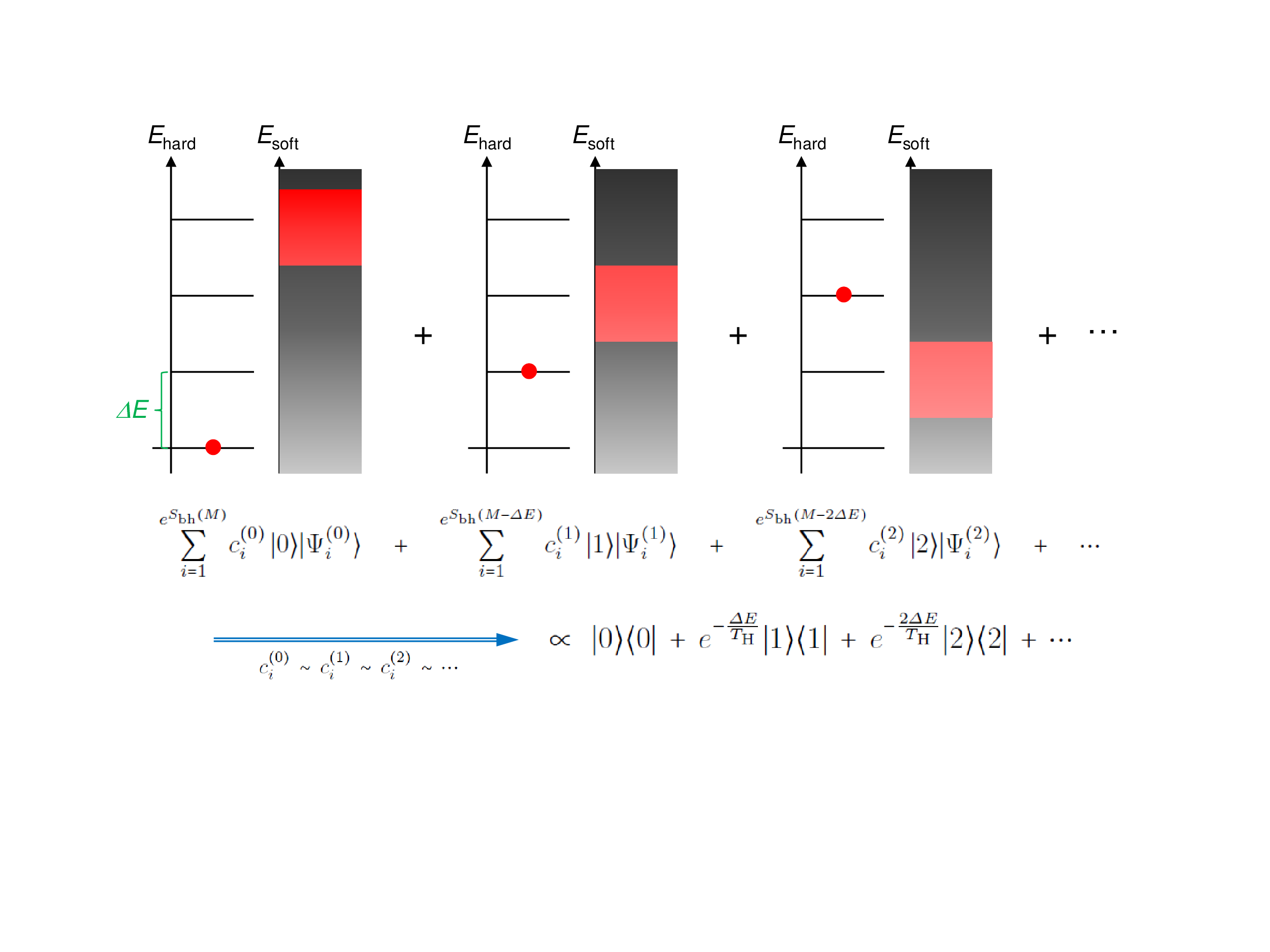}}
\caption{The entanglement between the hard and soft modes generated by the energy constraint is the origin of the thermality of a black hole at the semiclassical level.}
\label{fig:hard-soft}
\end{figure}

Note that the right-hand side of Eq.~(\ref{eq:sys-state}) contains terms with varying energies of the hard modes, $E_n$, whose differences are larger than $\Delta$.
The soft mode states that appear with $\ket{\{ n_\alpha \}}$, therefore, have energies $M - E_n$ (up to precision $\Delta$) to compensate the energy carried by the hard modes.
The number of independent such states are given by $S_{\rm bh}(M-E_n) = S_{\rm bh}(M) - E_n/T_{\rm H}$, which depends on $E_n$.
(The entropy density of soft modes is given by the Bekenstein-Hawking entropy density because most of the black hole entropy, as well as the energy, is carried by the soft modes; see Ref.~\refcite{Langhoff:2020jqa}.)
This is illustrated in Fig.~\ref{fig:hard-soft}.

The index $A$ of $\ket{\Psi_{A,0}(M)}$ in Eq.~(\ref{eq:sys-state}) labels microstates specified by the coefficients $c^A_{n i_n a}$.
The number of independent microstates $e^{S_{\rm tot}}$ is determined by the coarse-grained entropies of the soft modes $S_{\rm bh}(E_{\rm soft})$ and the far modes/early radiation $S_{\rm rad}$, and we let the index $A$ label the orthonormal basis states (of an arbitrary basis): $A = 1,\cdots,e^{S_{\rm tot}}$.
Here,%
\footnote{Note that $\ket{\Psi_{A,0}(M)}$ represent microstates of the soft modes and far modes (radiation) with the black hole put in the semiclassical vacuum, so that a generic state in the Hilbert space of dimension $e^{S_{\rm tot}}$ has the black hole of mass $M$.}
\begin{equation}
  e^{S_{\rm tot}} = \sum_n e^{S_{\rm bh}(M-E_n)} e^{S_{\rm rad}} = z\, e^{S_{\rm bh}(M)+S_{\rm rad}},
\end{equation}
where
\begin{equation}
  z \equiv \sum_n e^{-\frac{E_n}{T_{\rm H}}}.
\label{eq:def-z}
\end{equation}
With this convention, the coefficients $c^A_{n i_n a}$ satisfy 
\begin{equation}
  \sum_n \sum_{i_n = 1}^{e^{S_{\rm bh}(M-E_n)}} \sum_{a = 1}^{e^{S_{\rm rad}}} 
    c^{A*}_{n i_n a} c^{B}_{n i_n a} = \delta_{AB}.
\label{eq:norm}
\end{equation}

What else do we know about the coefficients $c^A_{n i_n a}$?
The spatial distribution of the soft modes is determined by the local Hawking temperature, Eq.~(\ref{eq:T_loc}).
This distribution is strongly peaked toward the stretched horizon, where the local temperature reaches the string scale.
Their internal dynamics, therefore, is controlled by the microscopic dynamics of quantum gravity and cannot be described by a low energy theory; indeed, we expect that it is nonlocal in the spatial directions along the horizon~\cite{Hayden:2007cs,Sekino:2008he}.
Nevertheless, it is widely believed that this dynamics exhibits certain characteristic behaviors:\ it is maximally quantum chaotic~\cite{Maldacena:2015waa}, fast scrambling~\cite{Hayden:2007cs,Sekino:2008he}, and does not have a feature discriminating low energy species beyond their spacetime and gauge properties~\cite{Banks:2010zn,Harlow:2018jwu}.

As discussed in Refs.~\refcite{Nomura:2018kia,Nomura:2019qps}, the dynamical properties described above are critical ingredients that distinguish the stretched horizon from normal material surfaces, leading to near empty interior spacetime.
Specifically, these properties imply that the coefficients $c^A_{n i_n a}$ in Eq.~(\ref{eq:sys-state}) have the statistical properties
\begin{equation}
  \vev{c^A_{n i_n a}} = 0,
\qquad
  \sqrt{\vev{|c^A_{n i_n a}|^2}} = \frac{1}{e^{\frac{1}{2}S_{\rm tot}}},
\label{eq:c-size}
\end{equation}
where $\vev{\cdots}$ represents an ensemble average over $(i_n,a)$, and that the phases of $c^A_{n i_n a}$'s are distributed uniformly.
Such a configuration is indeed reached quickly, within the scrambling time of order $(1/2\pi T_{\rm H})\ln S_{\rm bh}(M)$.
With Eq.~(\ref{eq:c-size}), we can trace out the soft modes, obtaining the thermal density matrix for the hard modes
\begin{equation}
  {\rm Tr}_{\rm soft} \ket{\Psi_{A,0}(M)} \bra{\Psi_{A,0}(M)}
  = \frac{1}{z} \sum_n e^{-\frac{E_n}{T_{\rm H}}} \ket{\{ n_\alpha \}} \bra{\{ n_\alpha \}} \otimes \rho_\phi,
\label{eq:rho_HR}
\end{equation}
where $\rho_\phi$ is an $n$-independent reduced density matrix for the far modes; fractional corrections to this expression are only of order $e^{-\frac{1}{2}S_{\rm bh}(M)}$.
This is the origin of the thermality of the black hole atmosphere in semiclassical theory~\cite{Nomura:2018kia}.
A black hole is a ``self thermalized'' system in which the degrees of freedom are confined near the stretched horizon due to its own gravitational potential.

Semiclassical excitations in the zone, i.e.\ hard mode excitations over a vacuum microstate, are described by annihilation and creation operators
\begin{align}
  b_\gamma &= \sum_n \sqrt{n_\gamma}\, 
    \ket{\{ n_\alpha - \delta_{\alpha\gamma} \}} \bra{\{ n_\alpha \}},
\label{eq:ann}\\*
  b_\gamma^\dagger &= \sum_n \sqrt{n_\gamma + 1}\, 
    \ket{\{ n_\alpha + \delta_{\alpha\gamma} \}} \bra{\{ n_\alpha \}}.
\label{eq:cre}
\end{align}
Since the semiclassical theory is not sensitive to the microstate of the black hole and is local in spacetime, these operators do not act on soft or far mode degrees of freedom.
Note that states obtained by acting these operators on $\ket{\Psi_{A,0}(M)}$ cannot be viewed as a vacuum state as they do not lead to the reduced density matrix of the form in Eq.~(\ref{eq:rho_HR}).%
\footnote{This fact, together with the fact that time of order the scrambling time is needed to reach the thermal equilibrium allows us to avoid the frozen vacuum problem of Ref.~\refcite{Bousso:2013ifa}; indeed, the effective theory of the interior (see below) erected after this time must see the semiclassical vacuum.~\cite{Hayden:2007cs}
 The Born rule problem of Ref.~\refcite{Marolf:2015dia} does not apply either due to the atypicality of the excited states.~\cite{Nomura:2019dlz}}

\begin{figure}[h!]
\centerline{\includegraphics[height=1.9in,trim = {0 0 0 0},clip]{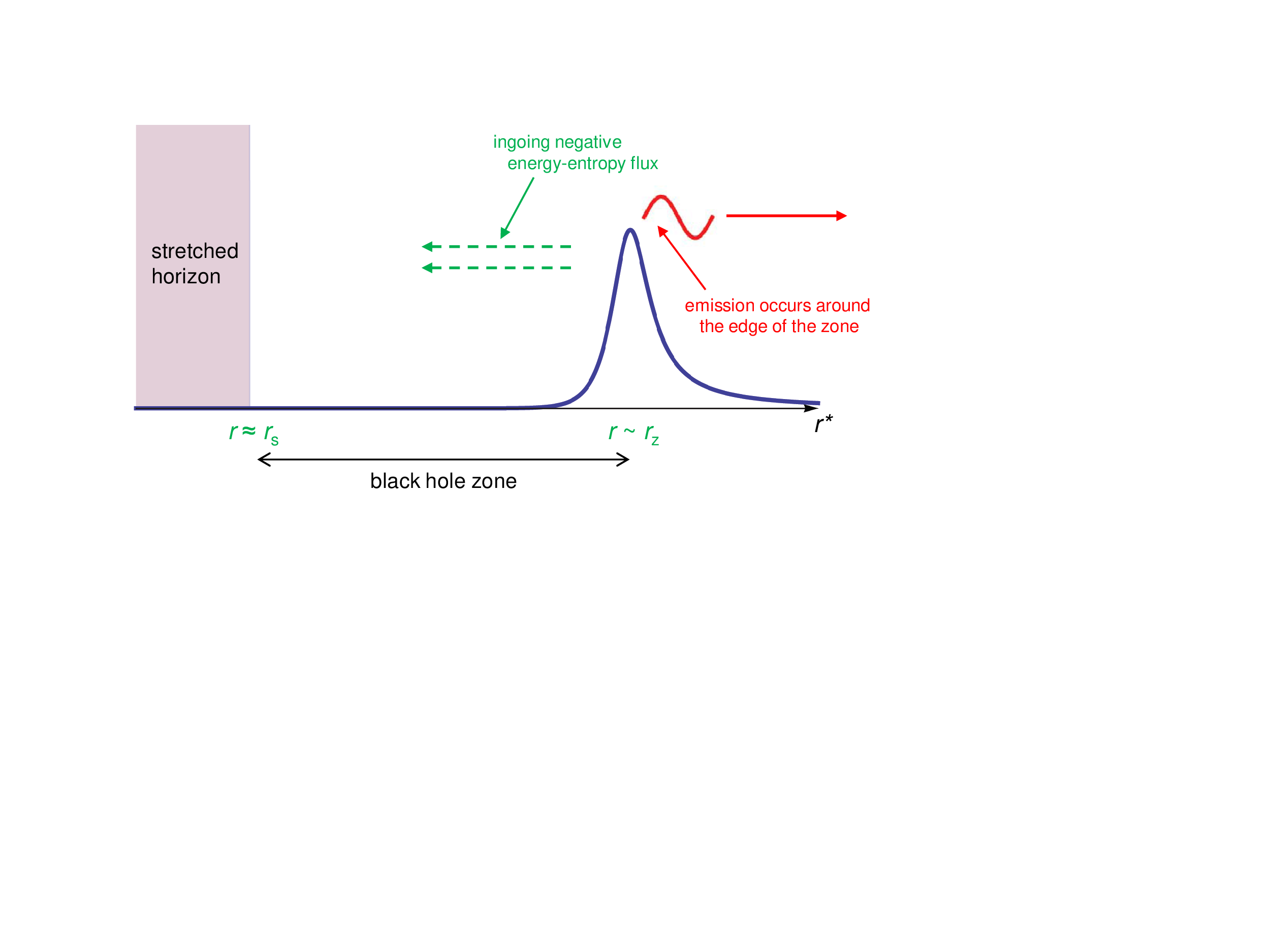}}
\caption{Hawking emission is viewed as occurring around the edge of the zone, where the information and energy of the black hole is extracted from the ($O(1)$ amount of) soft mode.
This produces an ingoing flux of negative energy and entropy, reducing the mass and entropy of the black hole.
Note that the horizontal axis is the tortoise coordinate $r^*$, in terms of which the wave length of a massless quantum does not change through propagation.}
\label{fig:emission}
\end{figure}

States described above evolve unitarily under boundary time evolution.
The time evolution of (a superposition of) microstates of the form of Eq.~(\ref{eq:sys-state})---particularly under the Hawking emission process---was discussed in Refs.~\refcite{Nomura:2018kia,Nomura:2014woa}.
While a complete description of the evolution requires a microscopic theory of quantum gravity, we can write down an evolution equation leaving the coefficients $c^A_{n i_n a}$ unspecified.
An important conclusion is that Hawking emission should be viewed as occurring through soft modes at the edge of the zone; see Fig.~\ref{fig:emission}.
This is possible because the distribution of the soft modes has a long tail in the zone, giving $O(1)$ degrees of freedom at the edge of the zone.
While this is fractionally a tiny amount of the whole black hole degrees of freedom, of order $1/S_{\rm bh}(M)$, the process is sufficient to carry away the energy and entropy of the black hole through its long lifetime.
(Similar pictures have also been discussed in Refs.~\refcite{Israel:2015ava,Giddings:2015uzr,Osuga:2016htn}.)

\subsection*{Effective theory of the interior}

As already emphasized a few times, the challenge of the unitary gauge description is to understand the black hole interior.
In particular, we want to understand what an object located in the zone and falling toward the black hole will experience after it crosses the horizon.
For this purpose, a description based on boundary time evolution is of little use.
In that description, the object will be absorbed into the stretched horizon when it gets there, after which no low energy description is available for it.
To describe the object's experience after reaching the horizon, we need a different time evolution associated with the proper time of the object.

Since all the fundamental degrees of freedom in the unitary gauge description exist outside the stretched horizon, we have to find the degrees of freedom that effectively describe the interior within these exterior degrees of freedom.
This boils down to identifying the degrees of freedom that can play the role of the ``second exterior'' of an analytically extended two-sided black hole~\cite{Papadodimas:2012aq}.
As discussed in Refs.~\refcite{Nomura:2018kia,Nomura:2019qps,Nomura:2020ska}, this can be done at each boundary time, and the degrees of freedom can be identified in the combined system of the soft and far modes.

Suppose that the state of the system at a boundary time $t$ is given by Eq.~(\ref{eq:sys-state}) (possibly) with excitations of hard modes and/or far modes over it.
We define normalized {\it mirror microstates} $\ketc{\{ n_\alpha \}_A}$ as the state of the soft and far modes entangled with the hard mode state $\ket{\{ n_\alpha \}}$ in the corresponding vacuum microstate:
\begin{equation}
  \ketc{\{ n_\alpha \}_A} = \alpha^A_n \sum_{i_n = 1}^{e^{S_{\rm bh}(M-E_n)}} \sum_{a = 1}^{e^{S_{\rm rad}}} c^A_{n i_n a} \ket{\psi^{(n)}_{i_n}} \ket{\phi_a}.
\label{eq:ketc}
\end{equation}
Here, the normalization constant $\alpha^A_n$ is given by
\begin{equation}
  \alpha^A_n = \frac{1}{\sqrt{\sum_{i_n = 1}^{e^{S_{\rm bh}(M-E_n)}} \sum_{a = 1}^{e^{S_{\rm rad}}} c^{A*}_{n i_n a} c^A_{n i_n a}}}
  = \sqrt{z}\,\, e^{\frac{E_n}{2T_{\rm H}}} \left[ 1 + O\bigl(e^{-\frac{1}{2}S_{\rm tot}}\bigr) \right],
\label{eq:alpha_nA}
\end{equation}
where $z$ is defined by Eq.~(\ref{eq:def-z}), and we have used statistical properties of $c^A_{n i_n a}$ to obtain the last expression.
The mirror microstates defined in this way are orthonormal up to exponentially small corrections:
\begin{equation}
  \innerc{\{ m_\alpha \}_A}{\{ n_\alpha \}_B} 
  = \delta_{m n}\, \delta_{AB} + O\bigl(e^{-\frac{1}{2}S_{\rm tot}}\bigr).
\label{eq:inner-c}
\end{equation}

With this definition, the vacuum microstate in Eq.~(\ref{eq:sys-state}) can be written as
\begin{equation}
  \ket{\Psi_{A,0}(M)} = \frac{1}{\sqrt{z}} \sum_n e^{-\frac{E_n}{2T_{\rm H}}} \ket{\{ n_\alpha \}} \ketc{\{ n_\alpha \}_A},
\label{eq:TFD}
\end{equation}
up to exponentially suppressed corrections.
This takes the form of the standard thermofield double state in the two-sided black hole picture,~\cite{Unruh:1976db,Israel:1976ur} so that we can define {\it canonical mirror operators} for microstate $A$ which play the role of the annihilation and creation operators in the zone of the second exterior region:
\begin{align}
  \tilde{b}_\gamma^A &= \sum_n \sqrt{n_\gamma}\, \Lketc{\{ n_\alpha - \delta_{\alpha\gamma} \}_A} \Lbrac{\{ n_\alpha \}_A},
\label{eq:ann-m-orig}\\*
  \tilde{b}_\gamma^{A\dagger} &= \sum_n \sqrt{n_\gamma + 1}\, \Lketc{\{ n_\alpha + \delta_{\alpha\gamma} \}_A} \Lbrac{\{ n_\alpha \}_A}.
\label{eq:cre-m-orig}
\end{align}
We can then define {\it infalling mode operators} for each microstate $A$:
\begin{align}
  a_\xi^A &= \sum_\gamma \bigl( \alpha_{\xi\gamma} b_\gamma + \beta_{\xi\gamma} b_\gamma^\dagger + \zeta_{\xi\gamma} \tilde{b}_\gamma^A + \eta_{\xi\gamma} \tilde{b}_\gamma^{A\dagger} \bigr),
\label{eq:a_xi}\\*
  a_\xi^{A\dagger} &= \sum_\gamma \bigl( \beta_{\xi\gamma}^* b_\gamma + \alpha_{\xi\gamma}^* b_\gamma^\dagger + \eta_{\xi\gamma}^* \tilde{b}_\gamma^A + \zeta_{\xi\gamma}^* \tilde{b}_\gamma^{A\dagger} \bigr),
\label{eq:a_xi-dag}
\end{align}
where $\xi$ is the label in which the frequency associated with boundary time $t$ is traded with that associated with infalling time $\tau$, and $\alpha_{\xi\gamma}$, $\beta_{\xi\gamma}$, $\zeta_{\xi\gamma}$, and $\eta_{\xi\gamma}$ are the Bogoliubov coefficients calculable using the standard field theory method.

The existence of operators in Eqs.~(\ref{eq:a_xi},~\ref{eq:a_xi-dag}) is essentially what we want, but it is not quite sufficient.~\cite{Papadodimas:2015jra,Nomura:2020ska}
In general, the state at a boundary time $t_*$ takes the form
\begin{equation}
  \ket{\Psi(t_*)} = \sum_{A=1}^{S_{\rm tot}} \sum_I d_{A I}(t_*) \ket{\Psi_{A,I}(M)},
\label{eq:Psi-t*}
\end{equation}
where $I$ labels excitations over the black hole background; specifically, we can imagine that they represent an infalling object in the zone.
Since this state does not generally factorize into a product of two states having the indices $A$ and $I$, respectively, the existence of infalling mode operators for a microstate is not enough to describe the fate of the state.

This problem is addressed by considering {\it globally promoted} mirror operators
\begin{equation}
  \tilde{\cal B}_\gamma = \sum_{A'=1}^{e^{S_{\rm eff}}} \tilde{b}_\gamma^{A'},
\qquad
  \tilde{\cal B}_\gamma^\dagger = \sum_{A'=1}^{e^{S_{\rm eff}}} \tilde{b}_\gamma^{A'\dagger},
\label{eq:global}
\end{equation}
where $\tilde{b}_\gamma^{A'}$ and $\tilde{b}_\gamma^{A'\dagger}$ are given by Eqs.~(\ref{eq:ann-m-orig},~\ref{eq:cre-m-orig}).
The corresponding infalling mode operators ${\cal A}_\xi$ and ${\cal A}_\xi^\dagger$ are given by the right-hand sides of Eqs.~(\ref{eq:a_xi},~\ref{eq:a_xi-dag}) with $\tilde{b}_\gamma^A$ and $\tilde{b}_\gamma^{A\dagger}$ replaced with $\tilde{\cal B}_\gamma$ and $\tilde{\cal B}_\gamma^\dagger$, respectively.
Here, $A'$ runs over orthonormal basis states in an $e^{S_{\rm eff}}$-dimensional subspace $\tilde{\cal M}$ of the vacuum microstate Hilbert space ${\cal M}$:
\begin{equation}
  \tilde{\cal M} = \Biggl\{ \sum_{A'=1}^{e^{S_{\rm eff}}} a_{A'} \ket{\Psi_{A',0}(M)} \,\Bigg|\, a_{A'} \in \mathbb{C},\, \sum_{A'=1}^{e^{S_{\rm eff}}} |a_{A'}|^2 = 1 \Biggr\},
\label{eq:tilde-cal-M}
\end{equation}
where $S_{\rm eff} < S_{\rm bh}(M) + S_{\rm rad}$.
As shown in Ref.~\refcite{Nomura:2020ska}, one can always choose $\tilde{\cal M}$ in such a way that the resulting infalling mode operators ${\cal A}_\xi$ and ${\cal A}_\xi^\dagger$ describe the fate of the state, Eq.~(\ref{eq:Psi-t*}).
In particular, the infalling Hamiltonian is given by
\begin{equation}
  H = \sum_\xi \Omega_\xi {\cal A}_\xi^\dagger {\cal A}_\xi + H_{\rm int}\bigl( \{ {\cal A}_\xi \}, \{ {\cal A}_\xi^\dagger \} \bigr),
\label{eq:tilde-H}
\end{equation}
where $\Omega_\xi$ is the frequency of mode $\xi$ with respect to infalling time $\tau$.

\begin{figure}[h!]
\centerline{\includegraphics[height=1.9in,trim = {0 0 0 0},clip]{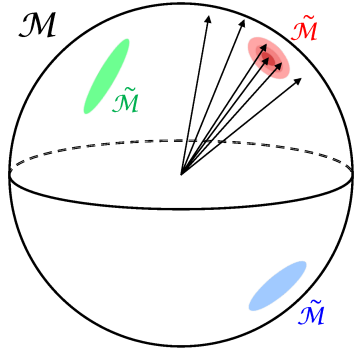}}
\caption{In the $e^{S_{\rm tot}}$-dimensional space ${\cal M}$ spanned by orthonormal vacuum microstates, we can build infalling mode operators that cover a subspace $\tilde{\cal M}$ in a state independent manner.
 The choice of $\tilde{\cal M}$ is arbitrary as represented by regions with different colors.
 Furthermore, the space $\tilde{\cal M}$ can be made larger as represented by the graded red regions, but only as long as ${\rm dim}\,\tilde{\cal M}$ is sufficiently smaller than ${\rm dim}\,{\cal M}$.
 This implies that a single $\tilde{\cal M}$ cannot cover a significant portion of ${\cal M}$.
 Note that the figure is only a schematic representation of the situation.}
\label{fig:M}
\end{figure}

Properties of global promotion and the resulting globally promoted operators have been studied in detail in Ref.~\refcite{Nomura:2020ska}.
Here we only list their salient features:
\begin{itemlist}
 \item 
 Infalling mode operators ${\cal A}_\xi$ and ${\cal A}_\xi^\dagger$ acting on vacuum microstates in $\tilde{\cal M}$ produce the same results (e.g.\ same correlators) as those in field theory, up to exponentially small errors of order
 \begin{equation}
   \epsilon = {\rm max}\left\{ \frac{e^{S_{\rm eff}}}{e^{S_{\rm bh}(M)+S_{\rm rad}}}, \frac{1}{e^{\frac{1}{2}S_{\rm bh}(M)+\frac{1}{2}S_{\rm rad}}} \right\}.
 \label{eq:epsilon}
 \end{equation}
 In fact, the construction of operators itself has an ambiguity of this order, which should be viewed as an intrinsic ambiguity of the semiclassical description.
 \item
 The vacuum microstate subspace in Eq.~(\ref{eq:tilde-cal-M}) can be taken as large as one wants unless the logarithm of its dimension, $S_{\rm eff}$, is exponentially close to $S_{\rm bh}(M) + S_{\rm rad}$.
 The operators ${\cal A}_\xi$, and ${\cal A}_\xi^\dagger$ then act only on the excitation index $I$, and not on the vacuum index $A'$, for any states built on a vacuum microstate in $\tilde{\cal M}$, up to corrections of order $\epsilon$.
 In other words, ignoring these exponentially small corrections, the Hilbert space can be viewed as
 \begin{equation}
   {\cal H} \approx {\cal H}_{\rm exc} \otimes ({\cal H}_{\rm vac} \cong \tilde{\cal M}),
 \label{eq:prod}
 \end{equation}
 where these mode operators act only on ${\cal H}_{\rm exc}$, which is the Hilbert space of the semiclassical theory.
 \item
 While $S_{\rm eff}$ can be taken to be a large fraction of $S_{\rm bh}(M) + S_{\rm rad}$:
 \begin{equation}
   S_{\rm eff} = c\, \{ S_{\rm bh}(M) + S_{\rm rad} \}
 \qquad
   (0 < c < 1),
 \label{eq:local-Seff}
 \end{equation}
 the fact that $c$ cannot be exponentially close to $1$ implies that we cannot make it sufficiently large to cover most of the states in ${\cal M}$, i.e.\ to make ${\cal A}_\xi$, and ${\cal A}_\xi^\dagger$ ``fully global'' state-independent operators.
 In fact, for $c > 1/2$, there is a simple relation between the size of $\tilde{\cal M}$ and the error $\epsilon$ for using operators promoted to cover $\tilde{\cal M}$:
 \begin{equation}
   \frac{{\rm dim}\, \tilde{\cal M}}{{\rm dim}\, {\cal M}} \,\approx\, \epsilon.
 \end{equation}
 Therefore, if we want to keep the error small, $\epsilon \ll 1$, the operators can be used only for a small fraction of states in ${\cal M}$.
 This is the statement of state dependence in Refs.~\refcite{Papadodimas:2013jku,Papadodimas:2015jra}.
 In fact, to cover all states in ${\cal M}$, we need double exponentially large number, $O(e^{e^{S_{\rm tot}-S_{\rm eff}}})$, of $\tilde{\cal M}$'s.
 This situation is illustrated in Fig.~\ref{fig:M}.
\end{itemlist}

\begin{figure}[h!]
\centerline{\includegraphics[height=1.9in,trim = {0 0 0 0},clip]{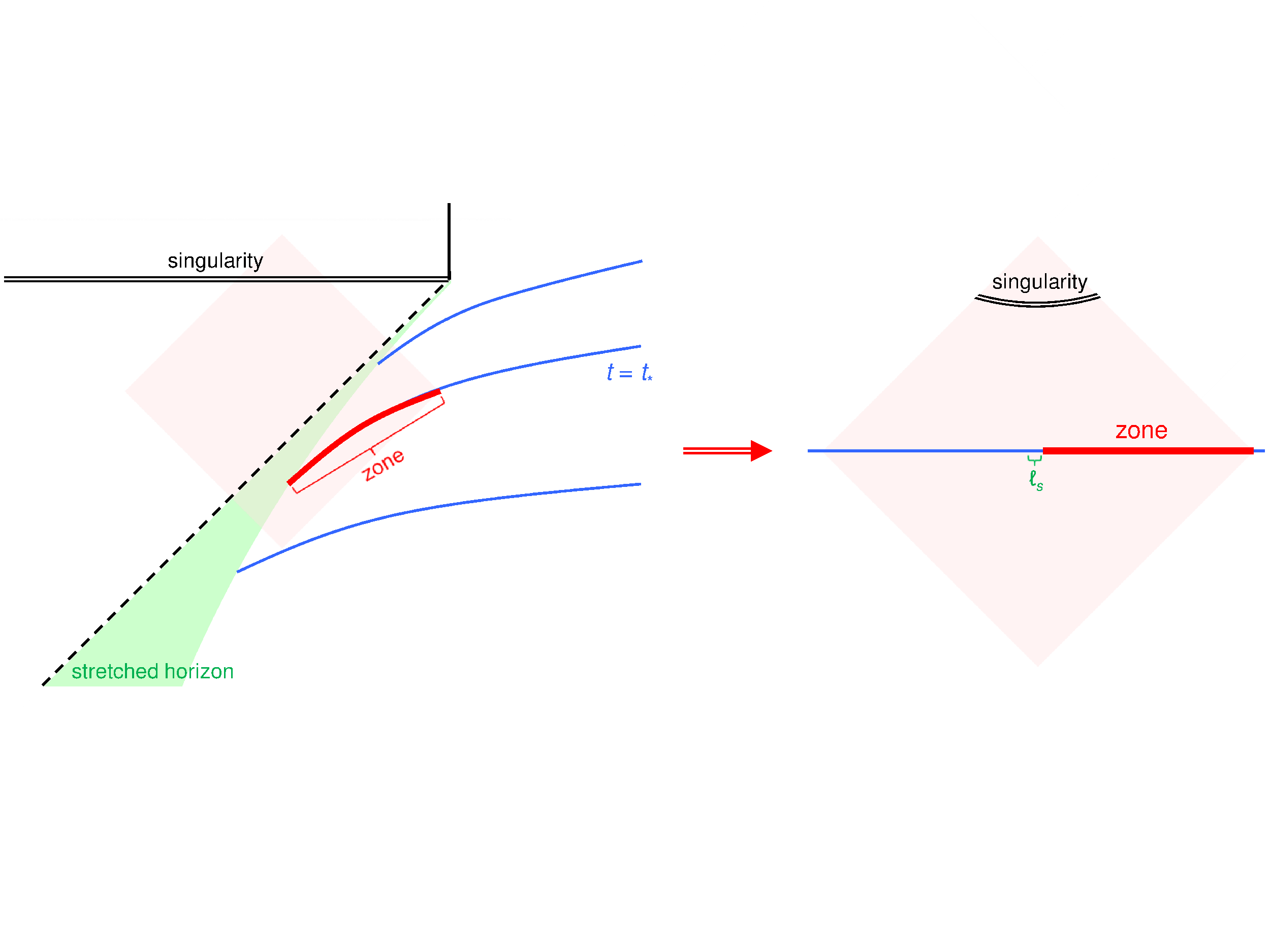}}
\caption{The effective theory of the interior can be erected at a boundary time $t_*$ in the original unitary theory having a one-sided black hole.
 It has an effective two-sided black hole geometry and describes physics in the causal domain of the union of the zone and its mirror region on the spatial hypersurface corresponding to the $t_*$ hypersurface.
 The effective theory is intrinsically semiclassical and cannot describe physics below the string length $l_{\rm s}$.}
\label{fig:two-sided}
\end{figure}

With the operators ${\cal A}_\xi$, and ${\cal A}_\xi^\dagger$, we can describe the interior of the black hole as in standard field theory.
Note that these operators are constructed out of annihilation and creation operators for the hard modes---semiclassical modes in the zone---as well as their mirrors in the second exterior of the effective two-sided geometry.
The state and operators of the effective theory, therefore, are defined on the union $U_0$ of the zone and its mirror region on the hypersurface of infalling time $\tau = 0$, which is matched to the boundary time $t_*$.
This implies that the theory describes only the limited spacetime region:\ the domain of dependence, $D(U_0)$, of $U_0$; see Fig.~\ref{fig:two-sided}.~\cite{Nomura:2018kia,Nomura:2019qps,Nomura:2020ska}
In order to cover a larger portion of the black hole interior, we must use multiple effective theories erected at different times.
This provides a specific realization of the idea of black hole complementarity.~\cite{Susskind:1993if,Susskind:1993mu}~%
\footnote{Unlike what is envisioned in some of the early works (e.g.\ Ref.~\refcite{Kiem:1995iy}), however, the exterior and interior descriptions are not related by a unitary transformation; rather, the latter emerges through coarse graining and is intrinsically semiclassical.}

The effective theory of the interior erected in this way is intrinsically semiclassical.
The picture of the interior emerges as a collective phenomenon of fundamental degrees of freedom, involving both the soft modes and far modes (early radiation).~\cite{Maldacena:2013xja,Nomura:2018kia,Nomura:2019qps}~%
\footnote{For a young black hole, i.e.\ a black hole with $S_{\rm rad} < S_{\rm bh}(M)$, the interior operators can be constructed only out of the soft modes, using the so-called Petz map.~\cite{Nomura:2020ska,Nomura:2019dlz}
 For an old black hole, this option is not available; the operators must involve both the soft and far modes.}
Being obtained through coarse graining, the theory is not fully unitary; particles can go outside $D(U_0)$ or, more importantly, hit the singularity.
In fact, this is a theory describing the dynamics of a finite number of degrees of freedom, which is the origin of the intrinsic ambiguity of order $\epsilon$ discussed earlier.

We emphasize that the construction of the interior works because of the special properties of the stretched horizon leading to Eq.~(\ref{eq:c-size}).
The criterion necessary for a surface to behave as a stretched horizon is stronger than that for regular thermalization occurring around us; in particular, it must exhibit ``universal thermalization'' applicable throughout all the low energy species.
Such a strong universality arises presumably only as a result of the string dynamics.
In the unitary gauge description, the formation of a black hole is signaled by the emergence of this UV dynamics, albeit in a highly redshifted form.

\subsection*{Apparent violation of the entropy bound, ensemble nature, etc}

\begin{figure}[h!]
\centerline{\includegraphics[height=1.8in,trim = {0 0 0 0},clip]{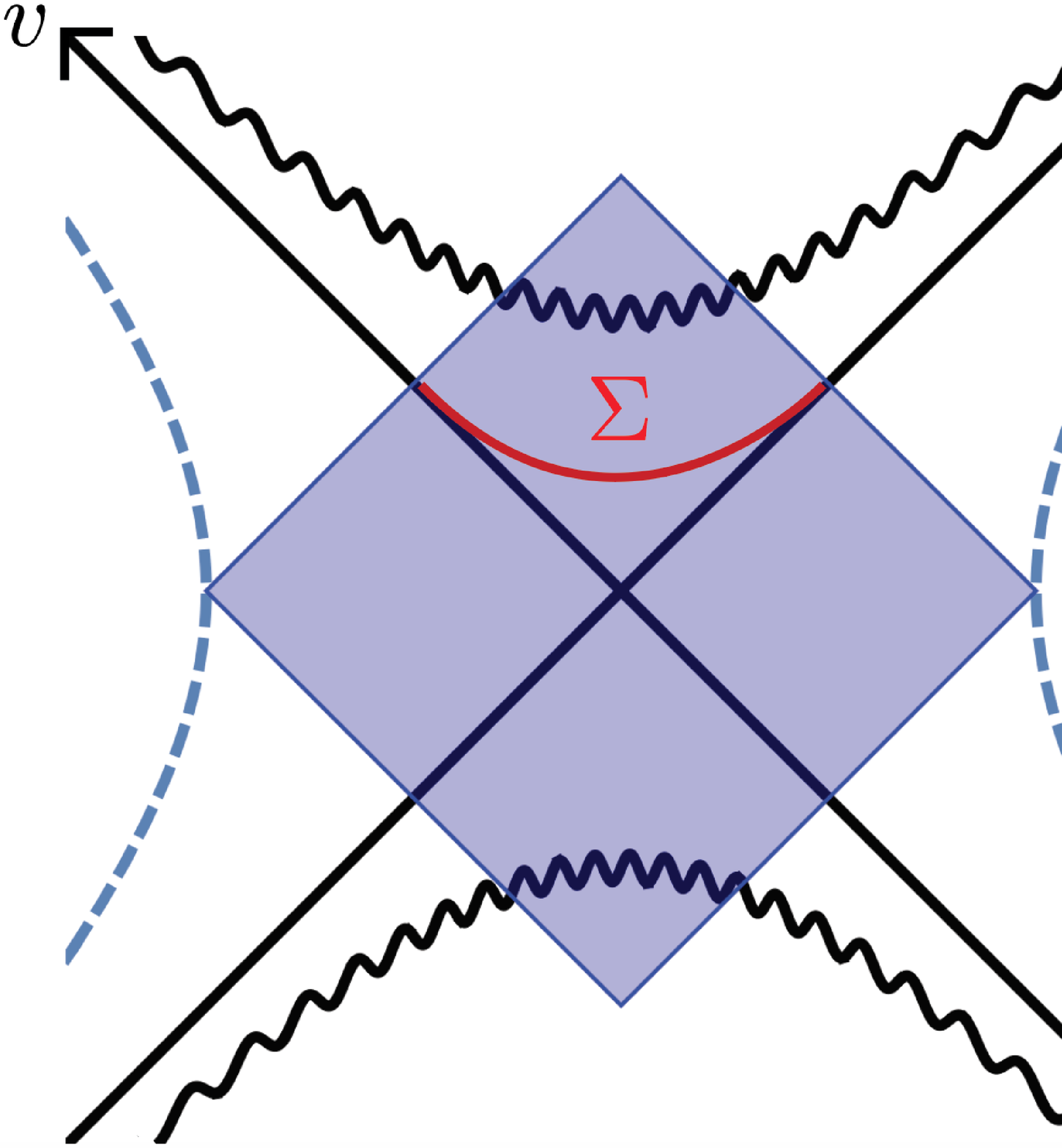}}
\caption{Within the spacetime region described by the effective theory of the interior (diamond at the center), the interior hypersurface having the maximal volume ($\Sigma$ in red) is bounded by the codimension-2 surfaces given by the intersections of the horizon and future-directed light rays emitted from $r = r_{\rm z}$ and its mirror.
 The volume of this hypersurface is finite.}
\label{fig:volume}
\end{figure}

The apparent violation of the Bekenstein-Hawking entropy bound encountered in the global spacetime description does not occur in the unitary gauge description.
In fact, the maximal interior volume one can consider in the effective theory is that of hypersurfaces bounded by the codimension-2 surfaces given by the intersections of the horizon and future-directed light rays emitted from $r = r_{\rm z}$ and its mirror; see Fig.~\ref{fig:volume}.
This volume is finite, and the amount of entropy of semiclassical matter one can place in this volume is indeed much smaller than the Bekenstein-Hawking entropy of the black hole.

The ensemble nature of the interior states observed in the global spacetime description is also reproduced.~\cite{Langhoff:2020jqa}
(See also Refs.~\refcite{Pollack:2020gfa,Belin:2020hea,Liu:2020jsv}.)
Consider an ensemble of soft mode (black hole) microstates defined by the collection of randomly selected states $\ket{\psi}_A$ ($A = 1,\cdots,{\cal K}$)
\begin{equation}
  \ket{\psi}_A = \sum_{i = 1}^{e^{S_{\rm bh}}} d^A_i \ket{\psi_i},
\qquad
  \sum_{i = 1}^{e^{S_{\rm bh}}} |d^A_i|^2 = 1,
\label{eq:soft-ens}
\end{equation}
where $\ket{\psi_i}$ ($i = 1,\cdots,e^{S_{\rm bh}}$) are the orthonormal basis states.
We first note that this ensemble can contain practically a double exponential number ${\cal N}_{\rm bh,eff}$ of ``independent'' states
\begin{equation}
  {\cal N}_{\rm bh,eff} \,\approx\, e^{e^{S_{\rm bh}}} \,\gg\, e^{S_{\rm bh}}.
\end{equation}
This can be seen by computing the inner product of two microstates $A \neq B$:
\begin{equation}
  {}_A\inner{\psi}{\psi}_B = \sum_{i = 1}^{e^{S_{\rm bh}}} d^{A*}_i d^B_i = O\bigl(e^{-\frac{S_0}{2}}\bigr),
\label{eq:inner}
\end{equation}
where we have used the fact that statistically $\sqrt{\vev{|d^A_i|^2}} \sim e^{-\frac{1}{2}S_{\rm bh}}$ and the phases of $d^A_i$ are distributed uniformly.
This is exponentially suppressed even if $\ket{\psi}_A$ and $\ket{\psi}_B$ are not orthogonal, unless $|(d^A_i - d^B_i)/d^A_i| \ll 1$ for the majority of $i$ (which requires a double exponential coincidence).
This is the origin of the apparent violation of the Bekenstein-Hawking entropy bound.
Note that ${}_A\inner{\psi}{\psi}_B$ in Eq.~(\ref{eq:inner}) have uniformly distributed random phases, reproducing the result of Eq.~(\ref{eq:ens-higherD}).

In order to apply a quasi-static description, we need to restrict our attention sufficiently small timescale $\varDelta t$.
A natural choice is $\varDelta t \lesssim \Delta$, in which case the soft modes cannot be resolved and hence should be integrated over.
The inner product of different interior states ($A \neq B$) is then obtained by averaging Eq.~(\ref{eq:inner}) over the space of microstates using the Haar measure:
\begin{equation}
  \overline{{}_A\inner{\psi}{\psi}_B } \,=\, \int\! dU\, {}_{U(A)}\inner{\psi}{\psi}_{U(B)} \,=\, \int\! dU \sum_{i = 1}^{e^{S_{\rm bh}}} d^{U(A)*}_i d^{U(B)}_i \,=\, 0,
\label{eq:inner-ave}
\end{equation}
where $\ket{\psi}_{U(A)}$ represents the state obtained by acting a unitary rotation $U$ on the state $\ket{\psi}_A$ in the space of microstates of dimension $e^{S_{\rm bh}}$.
Similarly, we find
\begin{equation}
  \overline{|\inner{\psi_A}{\psi_B}|^2} \,=\, \int\! dU \sum_{i,j = 1}^{e^{S_{\rm bh}}} d^{U(A)*}_i d^{U(B)}_i d^{U(B)*}_j d^{U(A)}_j \,=\, \delta_{AB} + O\bigl(e^{-S_{\rm bh}}\bigr)
\end{equation}
for general $A$ and $B$.
This gives Eq.~(\ref{eq:ens-lowerD}).

If the gravitational---bulk---theory is two dimensional, then there is no ensemble of soft modes, since the horizon is spatially a point.
In this case, however, the gravitational description emerges from an ensemble of unitary---boundary---theories.~\cite{Saad:2019lba,Stanford:2019vob}
Accordingly, a black hole state in the bulk corresponds to an ensemble of microstates in these unitary theories.
We thus find that the result in Eq.~(\ref{eq:ens-lowerD}) is also reproduced in this case.

\section{Beyond Black Holes}
\label{sec:beyond}

While the main focus of this article has been a black hole, many of the phenomena discussed are not specific to a black hole; rather, they are associated with the existence of a horizon.
We thus expect that a similar analysis applies to a cosmological horizon, including that of de~Sitter spacetime.
Relevant discussions include Refs.~\refcite{Nomura:2019qps,Nomura:2011dt} for the unitary gauge description and Refs.~\refcite{Chen:2020tes,Hartman:2020khs,Balasubramanian:2020xqf} for the global spacetime description.
Understanding this issue, indeed, seems to be vital in the cosmology of the eternally inflating multiverse.~\cite{Nomura:2011dt,Bousso:2011up,Nomura:2011rb}

\section{Discussion}
\label{sec:discuss}

Quantum gravity can be formulated in two different ways: through gravitational path integral and as a unitarily evolving holographic quantum system.
These two descriptions go well with the Lagrangian and Hamiltonian approaches, respectively, and as such make symmetries (general covariance) and unitarity manifest.
When there is a black hole, the starting points of the two descriptions are dramatically different, and yet they lead to the same physical conclusions.~\cite{Langhoff:2020jqa,Nomura:2020ska,Harlow:2020bee}

The coherence of the picture is rather convincing to conclude that a black hole has the smooth interior and yet evolves unitarily when viewed from the exterior.
An important point is that in both descriptions, the details of the microscopic dynamics are not necessary to reach the conclusion; certain basic assumptions are sufficient.
These assumptions, however, leave some questions.
\begin{itemlist}
 \item
 In the global spacetime description, what spacetime histories should we include in path integrals?
 What types of singularities, branch cuts, and so on should or should not be tolerated?
 \item
 In the unitary gauge description, what forces us to take particular, infalling mode operators to be the operators relevant to observables?
 The issue is presumably related to spacetime locality,~\cite{Nomura:2011rb,q-Darwinism,q-Darwinism-2} but the answer is not clear.
\end{itemlist}

Ultimately, these questions will be answered by the fundamental theory in a top-down manner, but from the low energy point of view, one can take unitarity and general covariance---or the equivalence principle allowing analytic extension of spacetime---as principles, or part of the definition of the theory.
With this, unitarity will instruct us whether we should or should not include a spacetime history in the global spacetime description, and general covariance requires us to take the infalling mode operators to be the operators relevant to classical observers in the unitary gauge description.
The fact that these two principles can be imposed at the same time, however, is still highly nontrivial as demonstrated by the historical struggles.~\cite{Hawking:1976ra,Almheiri:2012rt,Almheiri:2013hfa,Marolf:2013dba,Mathur:2009hf}

The perspective of quantum gravity developed here will have broader implications beyond black hole physics.
In particular, it will be important when the system develops a horizon, and as such it may play a vital role in understanding the cosmology of the eternally inflating multiverse.~\cite{Nomura:2011dt,Bousso:2011up,Nomura:2011rb}.

\section*{Acknowledgments}

I would like to thank Kevin Langhoff, Nico Salzetta, Jaime Varela, and Sean Weinberg for fruitful collaborations which have helped me to arrive at the picture presented here.
I would also like to thank Adam Bouland, Raphael Bousso, Chitraang Murdia, Masahiro Nozaki, Pratik Rath, and Arvin Shahbazi-Moghaddam for useful discussions throughout the course of exploring the subject.
This work was supported in part by the Department of Energy, Office of Science, Office of High Energy Physics under contract DE-AC02-05CH11231 and award DE-SC0019380.

\end{document}